\documentclass[preprint,review,12pt,3p]{elsarticle}

\usepackage{graphicx}
\usepackage{epsfig}
\usepackage{amssymb}
\usepackage{amsmath}
\usepackage{psfrag}
\usepackage{subfigure}
\usepackage{setspace}
\usepackage{booktabs}
\usepackage{multirow}

\usepackage{microtype}
\DisableLigatures{encoding = *, family = * }

\journal{Reliability Engineering \& System Safety}

\usepackage{hyperref}
\hypersetup{colorlinks=false}

\biboptions{sort&compress} 

\begin{document}


\setcounter{page}{1}
\begin{frontmatter}
\title{Quantitative model validation techniques: new insights}
\author{You Ling and Sankaran Mahadevan\corref{corauthor}}
\address{Department of Civil and Environmental Engineering, Vanderbilt University, TN 37235}
\cortext[corauthor]{Corresponding author\\Telephone: (615)322-3040, Email: sankaran.mahadevan@vanderbilt.edu}
\date{Dec 10, 2011}

\begin{abstract}
This paper develops new insights into quantitative methods for the validation of computational model prediction. Four types of methods are investigated, namely classical and Bayesian hypothesis testing, a reliability-based method, and an area metric-based method. Traditional Bayesian hypothesis testing is extended based on interval hypotheses on distribution parameters and equality hypotheses on probability distributions, in order to validate models with deterministic/stochastic output for given inputs. Formulations and implementation details are outlined for both equality and interval hypotheses. Two types of validation experiments are considered - fully characterized (all the model/experimental inputs are measured and reported as point values) and partially characterized (some of the model/experimental inputs are not measured or are reported as intervals). Bayesian hypothesis testing can minimize the risk in model selection by properly choosing the model acceptance threshold, and its results can be used in model averaging to avoid Type I/II errors. It is shown that Bayesian interval hypothesis testing, the reliability-based method, and the area metric-based method can account for the existence of directional bias, where the mean predictions of a numerical model may be consistently below or above the corresponding experimental observations. It is also found that under some specif{}ic conditions, the Bayes factor metric in Bayesian equality hypothesis testing and the reliability-based metric can both be mathematically related to the $p$-value metric in classical hypothesis testing. Numerical studies are conducted to apply the above validation methods to gas damping prediction for radio frequency (RF) microelectromechanical system (MEMS) switches. The model of interest is a general polynomial chaos (gPC) surrogate model constructed based on expensive runs of a physics-based simulation model, and validation data are collected from fully characterized experiments.
\end{abstract}

\begin{keyword}
model validation \sep hypothesis testing \sep Bayesian statistics \sep reliability \sep MEMS
\end{keyword}

\end{frontmatter}
\section{Introduction}\label{section:intro}
Model validation is def{}ined as the process of determining the degree to which a model is an accurate representation of the real world from the perspective of the intended use of the model~\cite{AIAA1998,ASME2006}. Qualitative validation methods such as graphical comparisons between model predictions and experimental data are widely used in engineering. However, statistics-based quantitative methods are needed to supplement subjective judgments and to systematically account for errors and uncertainty in both model prediction and experimental observation~\cite{Oberkampf2002}. 

Previous research ef{}forts include the application of statistical hypothesis testing methods in the context of model validation~\cite{Hartmann1995,Rebba2006,Rebba2006a,Rebba2008}, and development of validation metrics as measures of agreement between model prediction and experimental observation~\cite{Hills2003,Oberkampf2006,Rebba2008,Ferson2008,Ferson2009}. Some discussions on the pros and cons of these validation methods can be found in~\cite{Rebba2008,Liu2011a}. Based on these existing methods and the related studies, this paper is motivated by several issues which remain unclear in the practice of model validation: (1) validation with fully characterized vs. partially characterized experimental data; (2) validation of deterministic vs. stochastic model predictions; (3) accounting for the existence of directional bias; and (4) choice of thresholds in dif{}ferent validation metrics.

First, there are two possible types of validation data, resulting from (1) fully characterized experiments (i.e., all the inputs of the model/experiment are measured and reported as point values) or (2) partially characterized experiments (i.e., some inputs of the model/experiment are not measured or are reported as intervals). For instance, some input variables of the model/experiment may not be measured, but we may have expert opinions about the possible ranges or probability distributions of these input variables, and thus this experiment is "partially" characterized. In other words, there will be more uncertainty in the data from partially characterized experiments than from fully characterized experiments, due to the uncertainty in the input variables. Some partially characterized experiments with limited uncertainty may be considered for validation by practitioners. The term "input" is referred to as the variables in a model that can be measured in experiments. We assume that the same set of variables goes into the model and validation experiments as inputs, and we are comparing the outputs of the model and experiments during validation. Therefore, the terms "model inputs" and " experimental inputs" mean the same thing in this paper. When a model is developed, the physical quantity $Y$ is postulated to be a function of a set of variables $\{\boldsymbol{x},\boldsymbol{\theta}\}$. This function is not exactly known and hence is approximated using a model with output $Y_m$. $Y$ is observable through some experiments and $\boldsymbol{x}$ are the measurable inputs variables to the experiments. Note that $\boldsymbol{\theta}$ are the variables that cannot be measured in the experiments and are called as "model parameters". A simple example of the measurable experimental inputs is the amplitude of loading applied on a cantilever beam, while the deflection of the beam is the measured quantity. Also note that the diagnostic quality and the bias in experiments are not considered as "input". Instead, they are classified as components of the measurement uncertainty, which is represented by $\varepsilon_D$ in this paper.  While most of the previous studies only focus on validation with fully characterized experimental data, this paper explores the use of both types of data in various validation methods.

Second, due to the existence of aleatory and epistemic uncertainty, both the model prediction (denoted as $Y_m$) and the physical quantity to be predicted (denoted as $Y$) can be uncertain, and this has been the dominant case studied in the literature~\cite{Hills2003,Zhang2003a,Rebba2006,Rebba2006a,Rebba2008,Ferson2008,Ferson2009}. However, in practice it is possible that either $Y_m$ or $Y$ can be considered as deterministic. Note that $Y_m$ is deterministic means that for given values of the model input variables, the output prediction of the model is deterministic. The application of various validation methods to these dif{}ferent cases will be covered in this paper.

Third, in this study, we def{}ined two terms to characterize the dif{}ference between model prediction and validation data - bias and directional bias. Bias is defined as the difference between the mean value of model predictions and the statistical mean value of experiment observations, and the term "directional bias" means that the direction of bias remains unchanged as one varies the inputs of model and experiment. This paper explores various validation methods in order to account for the existence of the directional bias.

Fourth, although dif{}ferent validation metrics are usually developed to measure the agreement between model prediction and validation data from dif{}ferent perspectives, this paper shows that under certain conditions some of the validation metrics can be mathematically related. These relationships may help decision makers to select appropriate validation metrics and the corresponding model acceptance/rejection thresholds.

Various quantitative validation metrics, including the $p$-value in classical hypothesis testing~\cite{Schervish1996}, the Bayes factor in Bayesian hypothesis testing~\cite{O'Hagan1995}, a reliability-based metric~\cite{Rebba2008}, and an area-based metric~\cite{Ferson2008,Ferson2009}, are investigated in this paper. Based on the original def{}inition of Bayes factor, we formulate two types of Bayesian hypothesis testing, one on the accuracy of predicted mean and standard deviation of model prediction, and the other one on the entire predicted probability distribution of the model prediction. These two formulations of Bayesian hypothesis testing can be applied to both fully characterized and partially characterized experiments. The use of these two types of experimental data in the other validation methods is also investigated. The f{}irst formulation of Bayesian hypothesis testing, along with the modif{}ied reliability-based method and the area metric-based method, takes into account the existence of directional bias. The mathematical relationships among the metrics used in classical hypothesis testing, Bayesian hypothesis testing, and the reliability-based method are investigated.

Section~\ref{section:validationMethods} presents the general procedure of quantitative model validation in the presence of uncertainty. Section~\ref{section:binHT} and~\ref{section:nonHT} investigate the aforementioned model validation methods for (1) fully characterized and partially characterized experimental data, (2) application to the case when model prediction and the quantity to be predicted may or may not be uncertain, (3) sensitivity to the existence of the directional bias,  and (4) the mathematical relationships among some of these validation methods. A numerical example is presented in Section~\ref{section:numericalExample} to illustrate the validation of a MEMS switch damping model, which is a generalized polynomial chaos (gPC) surrogate model~\cite{Xiu2002} that has been constructed to predict the squeeze-f{}ilm damping coef{}f{}icient. The gPC model is used to replace the expensive micro-scale f{}luid simulation model and thus expedite the probabilistic analysis of the MEMS device.
\section{Quantitative validation of model prediction}\label{section:validationMethods}

Suppose a computational model is constructed to predict an unknown physical quantity. Quantitative model validation methods involve the comparison between model prediction and experimental observation. In this paper, we use the following notations \newline
\indent- $Y$ represents the "true value" of the system response \newline
\indent- $Y_m$ is the model prediction of this true response $Y$ \newline
\indent- $Y_D$ is the experimental observation of $Y$
 
The development of validation metrics is usually based on assumptions on $Y$, $Y_m$, and $Y_D$, and these assumptions relate to the various sources of uncertainty and the types of available validation data. In order to select appropriate validation methods, the f{}irst step is to identity the sources of uncertainty and the type of validation data.

As mentioned earlier, the available validation data can be from fully characterized or partially characterized experiments. In the case of fully characterized experiments, the model/experimental inputs $\boldsymbol{x}$ are measured and reported as point values. The true value of the physical quantity ($Y$) and the output of model ($Y_m$) corresponding to these measured values of $\boldsymbol{x}$ will be deterministic if there are no other uncertainty sources existing in the physical system and the model. Note that $Y$ and $Y_m$ can still be stochastic because of other uncertainty sources other than the input uncertainty. For example, the Young's modulus of a certain material can be stochastic due to variations in the material micro-structure, and the output of a regression model for given inputs is stochastic because of the random residual term. If the experiment is partially characterized, some of the inputs $\boldsymbol{x}$ are not measured or are reported as intervals, and thus the uncertainty in $\boldsymbol{x}$ should be considered. In the Bayesian approach, the lack of knowledge (epistemic uncertainty) about $\boldsymbol{x}$ is represented through a probability distribution (subjective probability). Then, since both $Y$ and $Y_m$ are considered as functions of $\boldsymbol{x}$, they also get treated through probability distributions. Non-probabilistic approaches have also been proposed to handle the epistemic uncertainty; in this paper, we only focus on probabilistic methods.

Note that $Y_D$ results from the addition of measurement uncertainty to the true value of the physical quantity $Y$, i.e., $Y_D = Y+\varepsilon_D$, where $\varepsilon_D$ represents measurement uncertainty. Hence, the uncertainty in the experimental observation ($Y_D$) can be split into two parts, the uncertainty in the physical system response ($Y$) and the measurement uncertainty in experiments ($\varepsilon_D$). It should be noted that experimental data with poor quality can hardly provide any useful information on the validity of a model. The discussions in this paper are restricted to the cases where uncertainty in data (due to the uncertainty in measuring experimental input and output variables) is limited.

 Table~\ref{table:validationApp} summarizes the applicability of the various validation methods investigated in this paper to the dif{}ferent scenarios discussed above, and more details will be presented in Sections~\ref{section:binHT} and~\ref{section:nonHT}.

\begin{table}[h!]
\begin{center}
\caption{Scenarios of validation and the corresponding methods}
\label{table:validationApp}
\begin{tabular}{l|lll}
\hline
\multirow{2}{*}{Experimental data} & \multirow{2}{3.5cm}{Quantity $Y$ \\(to be predicted)} & \multirow{2}{3cm}{Prediction $Y_m$ (from model)} & \multirow{2}{2.25cm}{Applicable methods} \\
 & & & \\
\hline
\multirow{3}{*}{Fully characterized} & Stochastic & Deterministic & 1,2,4\\
 & Deterministic & Stochastic & 1,2,4,5\\
 & Stochastic & Stochastic & 1,2,3,4,5\\
\hline
\multirow{2}{*}{Partially characterized} & \multirow{2}{*}{Stochastic} & \multirow{2}{*}{Stochastic} & \multirow{2}{*}{1,2,3,4,5}\\
 & & & \\
\hline
\multicolumn{4}{l}{\emph{Methods considered:}} \\
\multicolumn{4}{l}{1. Classical hypothesis testing} \\
\multicolumn{4}{l}{2. Bayesian interval hypothesis testing} \\
\multicolumn{4}{l}{3. Bayesian equality hypothesis testing} \\
\multicolumn{4}{l}{4. Reliability-based method} \\
\multicolumn{4}{l}{5. Area metric-based method} \\
\multicolumn{4}{l}{\emph{Note:} $Y_D$ is always treated as a random variable due to measurement uncertainty}
\end{tabular}
\end{center}
\end{table}

\begin{figure}[h!]
\begin{center}
\includegraphics[width=0.95\textwidth]{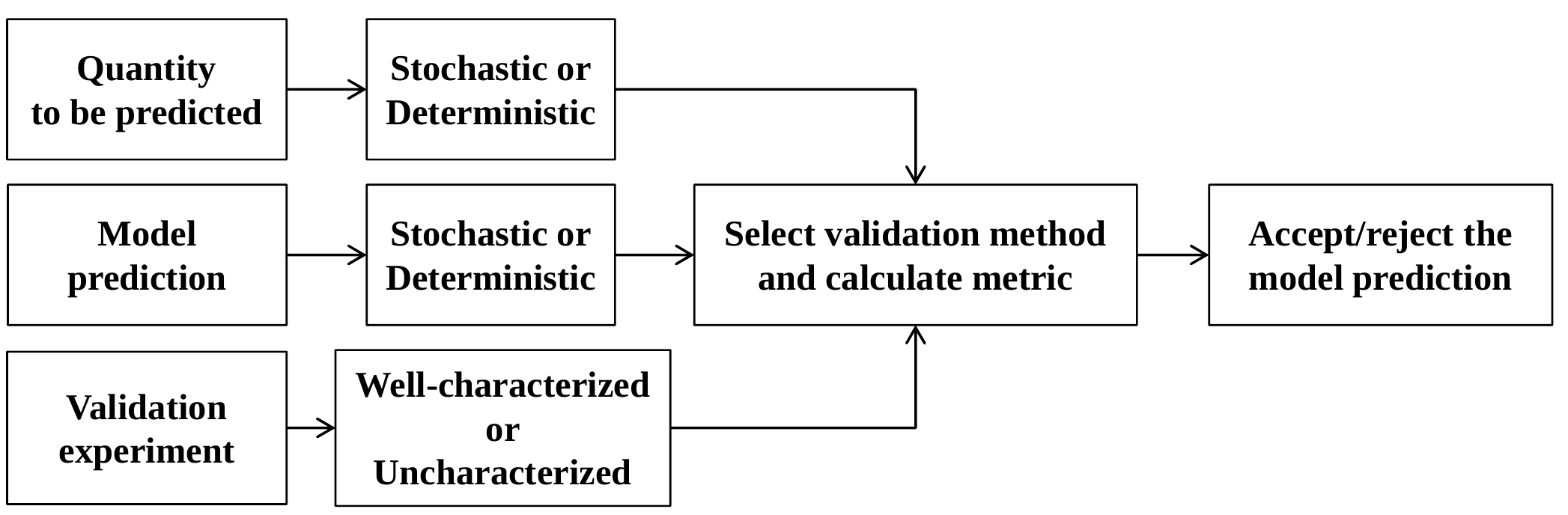}
\end{center}
\caption{Decision process in quantitative model validation}
\label{fig:flowchart}
\end{figure}

After selecting a validation method and computing the corresponding metric, another important aspect of model validation is to decide if one should accept or reject the model prediction based on the computed metric and the selected threshold. Section~\ref{section:binHT} and~\ref{section:nonHT} will provide some discussions on the decision threshold. The f{}lowchart in Fig.~\ref{fig:flowchart} describes a systematic procedure for quantitative model validation.

\section{Hypothesis testing-based methods}\label{section:binHT}
Statistical binary hypothesis testing involves deciding between the plausibility of two hypotheses - the null hypothesis (denoted as $H_0$) and the alternative hypothesis (denoted as $H_1$). $H_0$ is usually something that one believes could be true, whereas $H_1$ may be a rival of $H_0$~\cite{Marden2000}. For example, given a model for damping coef{}f{}icient prediction, $H_0$ can be the hypothesis that the model prediction is equal to the actual damping coef{}f{}icient, and correspondingly $H_1$ states that the model prediction is not equal to the actual damping coef{}f{}icient. The null hypothesis $H_0$ will be rejected if it fails the test, and will not be rejected if it passes the test. Two types of error can possibly occur from this exercise: rejecting the correct hypothesis (type I error), or failing to reject the incorrect hypothesis (type II error). In the context of model validation, it should be noted that the underlying subject matter domain knowledge is also necessary for the implementation of the hypothesis testing-based methods, especially in the formulation of test hypotheses ($H_0$ and $H_1$) and the selection of model acceptance threshold. To formulate appropriate $H_0$ and $H_1$ for the validation of a model with stochastic output prediction $Y_m$, we need to be clear about the physical interpretation of "model being correct". In other words, we need to decide whether or not the accurate prediction of certain order moments or the entire PDF of $Y_m$ suggests that the model is correct.

\subsection{Classical hypothesis testing}\label{section:CHT}
Classical hypothesis testing is well established and has been explained in detail in many statistics textbooks. A brief overview is given here, only to facilitate the development of mathematical relationships between various validation methods in later sections. 

In classical hypothesis testing, a test statistic is f{}irst formulated and the probability distributions of this statistic under the null and alternative hypothesis are derived theoretically or by approximations. Thereafter, one can compute the value of the test statistic based on validation data and thus calculate the corresponding $p$-value, which is the probability that the test statistic falls outside a range def{}ined by the computed value of the test statistic under the null hypothesis. The $p$-value can be considered as an indicator of how good the null hypothesis is, since a better $H_0$ corresponds to a narrower range def{}ined by the computed value of the test statistic and thus a higher probability of the test statistic falling outside the range. 

The practical outcome of model validation should be to provide useful information for decision making in terms of model selection. The decisions whether or not to reject the null hypothesis can be made based on the acceptable probabilities of making type I and type II errors (specified by the decision maker). The concept of signif{}icance level $\alpha$ def{}ines the maximum probability of making type I error, and the probability of making type II error $\beta$ can be estimated based on $\alpha$ and the probability distribution of the test statistic under $H_1$. The null hypothesis will be rejected if the computed $p$-value is less than $\alpha$, or the computed $\beta$ exceeds the acceptable value. Correspondingly, we will reject the model if $H_0$ is rejected, and accept the model if $H_0$ is not rejected. An alternative approach to comparing $p$-value and $\alpha$ is to use confidence intervals. A confidence interval can be constructed based on the confidence level $\gamma = 1 - \alpha$, and the null hypothesis will be rejected if the confidence interval does not include the predicted value from the model. 

It should be note that accepting the model (or failing to reject $H_0$) indicates that the accuracy of the model is acceptable, but it does not prove that the model (or $H_0$) is true. Also note that the comparison between $p$-value and signif{}icance level $\alpha$ becomes meaningless when the sample size of experimental data is large. Since almost no null hypothesis $H_0$ is true, the $p$-value will decrease as the sample size increases, and thus $H_0$ will tend to be rejected at a given signif{}icance level $\alpha$ as the sample size grows large~\cite{Marden2000}. In addition, the over-interpretation of $p$-values and the corresponding significance testing results can be misleading and dangerous for model validation. Criticisms on over-stressing $p$-values and significance levels can be found in~\cite{Ziliak2008,Ambaum2010}.

Various test statistics have been developed corresponding to dif{}ferent types of hypotheses. The $t$-test or $z$-test can be used to test the hypothesis that the mean of a normal random variable is equal to the model prediction; the chi-square test can be used to test the hypothesis that the variance of a normal random variable is equal to the model prediction; and the hypothesis that the observed data come from a specif{}ic probability distribution can be tested using methods such as the chi-square test, the Kolmogorov-Smirnov (K-S) test, the Anderson-Darling test and the Cramer test~\cite{Lehmann2005}. The tests on the variance or the probability distribution require relatively large amounts of validation data and thus only the tests on the distribution mean are discussed in this subsection, namely the $t$-test and the $z$-test.

The $t$-test is based on Student's $t$-distribution. Suppose the quantity to be predicted $Y$ is a normal random variable with mean $\mu$ and standard deviation $\sigma$, and the measurement noise $\varepsilon_D$ is also a normal random variable with zero mean and standard deviation $\sigma_D$. Thus, the experimental observation $Y_D = Y+\varepsilon_D \sim N(\mu,\sigma^2+\sigma_D^2)$. For the sake of simplicity, let $\sigma_{Y_D}=\sqrt{\sigma^2+\sigma_D^2}$. The validation data is a set of random samples of $Y_D$ with size $n$ (i.e., $y_{D1}$, $y_{D2}$, ..., $y_{Dn}$) and the corresponding sample mean is $\bar{Y}_D$ and sample standard deviation is $S_D$. The variable $(\bar{Y}_D-\mu)/(S_D/\sqrt{n})$ is modeled with a $t$-distribution with $(n-1)$ degrees of freedom. Therefore, if one assumes that the model mean prediction $\mu_m$ (if model prediction is deterministic, $\mu_m$ equals to the prediction value) is the mean of $Y$, i.e., the null hypothesis is $H_0: \mu = \mu_m$, then the corresponding test statistic $t$ (follows the same $t$-distribution) is
\begin{equation}\label{eq:tstatistic}
t = \frac{\bar{Y}_D-\mu_m}{S_D/\sqrt{n}}
\end{equation}

The $p$-value for the two-tailed $t$-test (considering both ends of the distribution) can be obtained as
\begin{equation}\label{eq:pvalueInttest}
p = 2F_{T,n-1}(-|t|)
\end{equation}
where $F_{T,n-1}$ is the cumulative distribution function (CDF) of a $t$-distribution with $(n-1)$ degrees of freedom. If the chosen signif{}icance level is $\alpha$, then one will reject the null hypothesis $H_0$ if  $p<a$, or fail to reject $H_0$ if $p>a$.

The $t$-test requires a sample size $n \geq 2$ in order to estimate the sample standard deviation $S_D$. If $n = 1$, the $z$-test can be used instead by assuming that the standard deviation of $Y$ is equal to the standard deviation of model prediction $Y_m$, i.e., $\sigma=\sigma_m$ and $\sigma_{Y_D}=\sqrt{\sigma_m^2+\sigma_D^2}$. Thus, the variable $(\bar{Y}_D-\mu)/(\sigma_{Y_D}/\sqrt{n})$ follows the standard normal distribution. Under the null hypothesis $H_0: \mu = \mu_m$, the test statistic $z$ is
\begin{equation}\label{eq:zstatistic}
z = \frac{\bar{Y}_D-\mu_m}{\sigma_{Y_D}/\sqrt{n}}
\end{equation}

The corresponding $p$-value for the two-tailed $z$-test can be computed as
\begin{equation}\label{eq:pvalueInztest}
p = 2\Phi(-|z|)
\end{equation}
where $\Phi$ is the CDF of the standard normal variable. Similar to the $t$-test, one will reject $H_0$ if $p<a$, or fail to reject $H_0$ if $p>a$.

To compute the probability of making type II error $\beta$, an alternative hypothesis $H_1$ is needed and a commonly seen formulation is $H_1: \mu = \mu_m + \epsilon_{\mu}$. In $t$-test, under the alternative hypothesis $H_1: \mu = \mu_m + \epsilon_{\mu}$, the $t$ statistic follows a non-central $t$-distribution with noncentrality parameter $\delta=\sqrt{n}\epsilon_{\mu}/\sigma_{Y_D}$~\cite{Srivastava2002,Mcfarland2008}, the probability of making type II error $\beta$ can then be estimated as
\begin{equation}
\beta = 1-Pr(|t| > t_{1-\alpha/2,n-1} | \delta)
\end{equation}
where the term $Pr(|t| > t_{\alpha/2,n-1} | \delta)$ is called the power of the test in rejecting $H_0$. Similarly, $\beta$ in the $z$-test can be estimated as
\begin{equation}
\beta = 1-Pr \big( |z-\delta| > \Phi^{-1}(1-\alpha/2) \big)
\end{equation}

Note that the above discussion is for the case that both $Y$ and $Y_m$ are stochastic. If $Y$ is deterministic, the standard deviation $\sigma$ becomes zero; if $Y_m$ is deterministic, $\sigma_m$ becomes zero. However, the computation procedure of $p$-value remains the same.

Applying classical hypothesis testing to fully characterized experiments is straightforward as one can directly compare the data with the model predictions for given inputs. For partially characterized experiments, some of the inputs of the model/experiments are available in the form of intervals or probability distributions based on measurements or expert opinions. Let data that have inputs with the same intervals or probability distributions form a sample set. The aforementioned $t$-test and $z$-test can then be conducted by comparing the mean of the sample set with the mean of the unconditional probability distribution of model output ("unconditional" means that the probability distribution is not dependent on the point values of inputs). The unconditional probability distribution of model output can be obtained by propagating uncertainty from the input variables to the output variable~\cite{Hills1999}.

\subsection{Bayesian hypothesis testing}\label{section:BHT}
In probability theory, Bayes' theorem reveals the relation between two conditional probabilities, e.g., the probability of occurrence of an event $A$ given the occurrence of an event $E$ (denoted as $Pr(A|E)$), and the probability of occurrence of the event $E$ given the occurrence of the event $A$ (denoted as $Pr(E|A)$). This relation can be written as~\cite{Haldar2000} 
\begin{equation}
Pr(E|A)=\frac{Pr(A|E)Pr(E)}{Pr(A)}
\end{equation}

Suppose event $A$ is the observation of a single validation data point $y_D$, and event $E$  can be either the hypothesis $H_0$ is true or the hypothesis $H_1$ is true. Using Bayes' theorem, we can calculate the ratio between the posterior probabilities of the two hypotheses given validation data $y_D$ as
\begin{equation}\label{eq:BHTformula}
\frac{Pr(H_0|y_D)}{Pr(H_1|y_D)} = \frac{Pr(y_D|H_0)}{Pr(y_D|H_1)}*\frac{Pr(H_0)}{Pr(H_1)}
\end{equation}
where $Pr(H_0)$ and $Pr(H_1)$ are the prior probabilities of $H_0$ and $H_1$ respectively, representing the prior knowledge one has on the validity of these two hypotheses before collecting experimental data; and $Pr(H_0 |y_D)$ and $Pr(H_1 |y_D)$ are the posterior probabilities of $H_0$ and $H_1$ respectively, representing the updated knowledge one has after analyzing the collected experimental data. The likelihood function $Pr(y_D|H_i)$ in Eq.~\ref{eq:BHTformula} is the conditional probability of observing the data $y_D$ given the hypothesis $H_i$ ($i = 0$ or $1$), and the ratio $Pr(y_D|H_0)/Pr(y_D|H_1)$ is known as the Bayes factor~\cite{O'Hagan1995,Kass1995} and is used as the validation metric.

The original intent of the Bayes factor was to compare the data support for two models~\cite{Pericchi2005}, and thus the two hypotheses become $H_0$: model $M_i$ is true and $H_1$: model $M_j$ is true. If $\boldsymbol{\theta_i}$ and $\boldsymbol{\theta_j}$ are the parameters of the two competing models respectively, the Bayes factor is calculated as
\begin{equation}\label{eq:DefBayesf}
B = \frac{Pr(y_D|H_0)}{Pr(y_D|H_1)} = \frac{\int{Pr(y_D|\boldsymbol{\theta_i})\pi(\boldsymbol{\theta_i})d\boldsymbol{\theta_i}}}{\int{Pr(y_D|\boldsymbol{\theta_j})\pi(\boldsymbol{\theta_j})d\boldsymbol{\theta_j}}}
\end{equation}
where $\pi(\boldsymbol{\theta_i})$ and $\pi(\boldsymbol{\theta_j})$ are the probability density distributions of $\boldsymbol{\theta_i}$ and $\boldsymbol{\theta_j}$ respectively.

In the context of validating a single model, $H_0$ and $H_1$ need to be formulated dif{}ferently. Rebba and Mahadevan~\cite{Rebba2008,Rebba2006} proposed the equality-based formulation ($H_0: y_m = y_D, H_1: y_m \neq y_D$) and the interval-based formulation ($H_0: |y_m - y_D|<\epsilon, H_1: |y_m - y_D|>\epsilon$) for Bayesian hypothesis testing, where $y_m$ is the model prediction for a particular input $\boldsymbol{x}$. 

Consider the case when both the model prediction $Y_m$ and the quantity to be predicted $Y$ are random variables. Two null hypotheses can be formulated: (1) the hypothesis that the dif{}ference between the means of $Y_m$ and $Y$, and the dif{}ference between the standard deviations of $Y_m$ and $Y$, are within desired intervals respectively; (2) the hypothesis that the PDF of $Y_m$ is equal to the PDF of $Y$. With the f{}irst formulation, it is straightforward to derive the likelihood functions under the null and alternative hypothesis, and the existence of directional bias can be ref{}lected in the test, as will be shown below. The advantages of the second formulation are that it avoids the setting of interval width in the f{}irst formulation, and leads to a direct test on probability distributions instead of distribution parameters. For the case that either $Y$ or $Y_m$ is deterministic, the f{}irst formulation can still be applicable by setting the standard deviation of the deterministic quantity to be zero; however, the second formulation only applies to the case when both $Y$ and $Y_m$ are stochastic. These two formulations are applicable to both fully characterized and partially characterized experiments. Note that in the case where the model output follows a tail-heavy distribution, formulating hypotheses on higher order moments (instead of the mean and standard deviation) may be necessary in order to assess the validity of the model. In this paper, the prediction distribution of the damping model in the application example is close to a Gaussian distribution. Therefore, we only consider hypotheses on the first two moments (mean and standard deviation) and the entire PDF for the purpose of illustration.

\paragraph{Interval hypothesis on distribution parameters}

The interval hypothesis can be formulated as $H_0: \epsilon_{\mu1} \le \mu_m-\mu \le \epsilon_{\mu2}, \epsilon_{\sigma1} \le \sigma_m-\sigma \le \epsilon_{\sigma2}$, and  $H_1: \mu_m-\mu>\epsilon_{\mu2} \text{ or } \mu_m-\mu<\epsilon_{\mu1}, \sigma_m-\sigma>\epsilon_{\sigma2} \text{ or } \sigma_m-\sigma<\epsilon_{\sigma1}$. $\mu_m$ and $\mu$ are the means of $Y_m$ and $Y$ respectively, and $\sigma_m$ and $\sigma$ are the standard deviations of $Y_m$ and $Y$ respectively. $\epsilon_{\mu1}$, $\epsilon_{\mu2}$, $\epsilon_{\sigma1}$ and $\epsilon_{\sigma2}$ are constants which def{}ine the width of interval. Note that $\epsilon_{\mu1}< \epsilon_{\mu2}$, $\epsilon_{\sigma1} < \epsilon_{\sigma2}$.

Under the interval hypothesis $H_0$, $\mu$ can be any value between [$\mu_m-\epsilon_{\mu2}$, $\mu_m-\epsilon_{\mu1}$]. So $\mu \sim \textnormal{Unif}(\mu_m-\epsilon_{\mu2}, \mu_m-\epsilon_{\mu1})$, and the PDF $\pi_0(\mu|\mu_m) = 1/(\epsilon_{\mu2}-\epsilon_{\mu1})$. Similarly, $\sigma \sim \textnormal{Unif} (\sigma_m-\epsilon_{\sigma2}, \sigma_m-\epsilon_{\sigma1})$, and the PDF $\pi_0(\sigma|\sigma_m) = 1/(\epsilon_{\sigma2}-\epsilon_{\sigma1})$. Thus
\begin{eqnarray}\label{eq:numIntBHT}
\pi_0(y|\mu_m,\sigma_m) &=& \int \int \pi(y|\mu,\sigma)\pi_0(\mu|\mu_m)\pi_0(\sigma|\sigma_m)d\mu d\sigma \nonumber\\
&=& \frac{1}{(\epsilon_{\mu2}-\epsilon_{\mu1})(\epsilon_{\sigma2}-\epsilon_{\sigma1})} \int_{\sigma_m-\epsilon_{\sigma2}}^{\sigma_m-\epsilon_{\sigma1}} \Big\{\int_{\mu_m-\epsilon_{\mu2}}^{\mu_m-\epsilon_{\mu1}} \pi(y|\mu,\sigma)d\mu\Big\} d\sigma
\end{eqnarray}

In the presence of measurement noise, the experimental observation is a random variable with conditional probability $Pr(y_D|y)$. Hence the likelihood function under the null hypothesis $H_0$ can be derived as
\begin{eqnarray}\label{eq:LKintBHT0}
Pr(y_D|H_0) &=& \int Pr(y_D|y)\pi_0(y|\mu_m,\sigma_m)dy
\end{eqnarray}

Under the alternative hypothesis $H_1$, $\mu$ can be any value outside [$\mu_m-\epsilon_{\mu2}$, $\mu_m-\epsilon_{\mu1}$], but the uniform distribution is not applicable to inf{}inite space in practical cases. To avoid this issue, we can assume that the possible values of $\mu$ are within a f{}inite interval [$\mu_l$, $\mu_u$] based on the underlying physics. Therefore $\mu \sim \textnormal{Unif}(\mu_l, \mu_m-\epsilon_{\mu2}) \cup (\mu_m-\epsilon_{\mu1},\mu_u)$, and the PDF $\pi_1(\mu|\mu_m) = 1/(\mu_u-\mu_l+\epsilon_{\mu1}-\epsilon_{\mu2})$. Similarly, $\sigma \sim \textnormal{Unif} (\sigma_l, \sigma_m-\epsilon_{\sigma2}) \cup (\sigma_m-\epsilon_{\sigma1},\sigma_u)$, and the PDF $\pi_1(\sigma|\sigma_m) =1/(\sigma_u-\sigma_l+\epsilon_{\sigma1}-\epsilon_{\sigma2})$. thus
\begin{eqnarray}\label{eq:denIntBHT}
\pi_1(y|\mu_m,\sigma_m) &=& \int \int \pi(y|\mu,\sigma)\pi_1(\mu|\mu_m)\pi_1(\sigma|\sigma_m)d\mu d\sigma \nonumber\\
&=&  \frac{A}{(\mu_u-\mu_l+\epsilon_{\mu1}-\epsilon_{\mu2})(\sigma_u-\sigma_l+\epsilon_{\sigma1}-\epsilon_{\sigma2})}
\end{eqnarray}
where A is calculated as
\begin{eqnarray}
A &=& \int_{\sigma_l}^{\sigma_m-\epsilon_{\sigma2}} \Big\{\int_{\mu_l}^{\mu_m-\epsilon_{\mu2}} \pi(y|\mu,\sigma)d\mu + \int_{\mu_m-\epsilon_{\mu1}}^{\mu_u} \pi(y|\mu,\sigma)d\mu\Big\} d\sigma + \nonumber\\
&& \int_{\sigma_m-\epsilon_{\sigma1}}^{\sigma_u} \Big\{\int_{\mu_l}^{\mu_m-\epsilon_{\mu2}} \pi(y|\mu,\sigma)d\mu + \int_{\mu_m-\epsilon_{\mu1}}^{\mu_u} \pi(y|\mu,\sigma)d\mu\Big\} d\sigma
\end{eqnarray}

The likelihood function under $H_1$ can then be derived as
\begin{eqnarray}\label{eq:LKintBHT1}
Pr(y_D|H_1) &=& \int Pr(y_D|y)\pi_1(y|\mu_m,\sigma_m)dy
\end{eqnarray}

The Bayes factor for the Bayesian interval hypothesis testing can be calculated by dividing $Pr(y_D|H_0)$ in Eq.~\ref{eq:LKintBHT0} by $Pr(y_D|H_1)$ in Eq.~\ref{eq:LKintBHT1}.

It is straightforward to apply this method to the case that $Y_m$ is deterministic and the case that $Y$ is deterministic. For the f{}irst case, let $\sigma_m$ be zero and the rest of the computation remains the same. For the second case, the interval assumption will only be made on $\mu$ and $\mu_m$, since we know $\sigma$ is zero. The other steps will be the same as above.

\begin{figure}[h!]
\begin{center}
\includegraphics[width=0.35\textwidth]{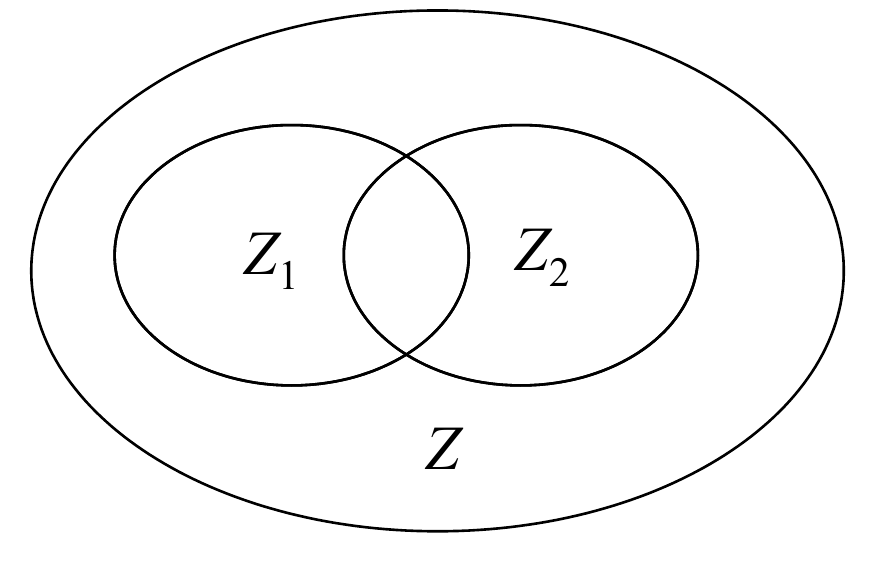}
\end{center}
\caption{Graphical illustration of the combined test}
\label{fig:setGraph}
\end{figure}

The directional bias def{}ined in Section~\ref{section:intro} can be captured by conducting two separate Bayesian interval hypothesis tests. In the f{}irst test, we set $\epsilon_{\mu1} = -\epsilon_{\mu}$ and $\epsilon_{\mu2} = 0$, and thus under the null hypothesis $-\epsilon_{\mu} \le \mu_m-\mu \le 0$. In the second test, we set $\epsilon_{\mu1} = 0$ and $\epsilon_{\mu2} = \epsilon_{\mu}$, and thus under the null hypothesis $0 \le \mu_m-\mu \le \epsilon_{\mu}$. The model will fail if any of these two null hypotheses fails the corresponding test. Therefore, the existence of directional bias will increase the chance of a model to fail the combined test. Fig.~\ref{fig:setGraph} illustrates this combined test using the concept of data space. Suppose $Z$ is the overall validation data space, $Z_1$ is the set of data which does not support the model in the f{}irst Bayesian interval hypothesis test, and $Z_2$ is the set of data which does not support the model in the second test. Then, the union of $Z_1$ and $Z_2$ is the set of data that does not support the model combining these two tests.

\paragraph{Equality hypothesis on probability density functions}

To further validate the entire distribution of $Y_m$ predicted by a probabilistic model, $H_0$ or $H_1$ can be formulated correspondingly as the predicted distribution $\pi(y_m)$ being or not being the true distribution of the quantity to be predicted $Y$, i.e., $H_0: \pi(y_m)=\pi(y)$, and $H_1: \pi(y_m) \neq \pi(y)$. The Bayes factor in this case becomes
\begin{equation}\label{eq:BFvalidation0}
B = \frac{Pr(y_D|H_0)}{Pr(y_D|H_1)} = \frac{\int{Pr(y_D|y)\pi_0(y)dy}}{\int{Pr(y_D|y)\pi_1(y)dy}}
\end{equation}
where $Pr(y_D|y)$ is the conditional probability of observing noisy data $y_D$ given that the actual value of $Y$ is $y$; $\pi_0(y)$ is the PDF of $Y$ under the null hypothesis $H_0$ and hence $\pi_0(y)=\pi(y_m)$; $\pi_1(y)$ is the PDF of $Y$ under the alternative hypothesis $H_1$. If no extra information about $\pi_1(y)$ is available, it can be assumed as a non-informative uniform PDF. Note that the bounds of this uniform distribution will af{}fect the value of the estimated Bayes factor, and thus it should be carefully selected according to the available information. 

Note that $Pr(y_D|y)$ is proportional to the value of the PDF of $Y_D$ conditioned on $y$ which is evaluated at $Y_D=y$, i.e., $Pr(y_D|y) \propto \pi(y_D|y)$. Therefore, Eq.~\ref{eq:BFvalidation0} can be rewritten as
\begin{equation}\label{eq:BFvalidation}
B = \frac{\int{\pi(y_D|y)\pi_0(y)dy}}{\int{\pi(y_D|y)\pi_1(y)dy}}
\end{equation}

\paragraph{Validation data from fully/partially characterized experiments}
If the validation data is from a fully characterized experiment, i.e., all the input parameters $\boldsymbol{x}$ of the experiment are  measured and the point values of $\boldsymbol{x}$ are available, $\mu_m$ and $\sigma_m$ used in the Bayesian interval hypothesis testing are the mean and standard deviation of the model prediction given the input $\boldsymbol{x}$, and the PDF of $Y_m$ ($\pi(y_m)$) used in the Bayesian equality hypothesis testing is conditioned on $\boldsymbol{x}$. If the experiment is partially characterized, i.e., some of input $\boldsymbol{x}$ corresponding to observation $y_D$ are not measured or are reported as intervals, we can assume that $\boldsymbol{x}$ have the PDF $\pi(\boldsymbol{x})$ based on reported intervals or expert opinions. One can f{}irst calculate the unconditional PDF of model prediction $\pi(y_m)$ via propagating uncertainty from $\boldsymbol{x}$ to model output $Y_m$
\begin{equation}
\pi(y_m) = \int{\pi(y_m|\boldsymbol{x})\pi(\boldsymbol{x})d\boldsymbol{x}}
\end{equation}
and then calculate $\mu_m$ and $\sigma_m$ from $\pi(y_m)$.
If data from both fully characterized and partially characterized experiments are available, we can f{}irst calculated Bayes factors corresponding to these two types of data points separately using dif{}ferent $\mu_m$ and $\sigma_m$ (in the Bayesian interval hypothesis testing), or $\pi(y_m)$ (in the Bayesian equality hypothesis testing) as shown above, and then multiply these Bayes factors to obtain an overall Bayes factor, as discussed below.

\paragraph{Bayesian hypothesis testing with multiple data points}
If there are $N$ ($N>1$) validation data points available and the corresponding experiments are conducted independently, i.e., no correlation exists between dif{}ferent data points, according to the basic rule of probability theory, the probability of observing the whole data set $Pr(y_D)$ is the product of the probabilities of observing individual data points $Pr(y_{Di})$, $i = 1,2,...,N$. Since the likelihood functions $Pr(y_D|H_0)$ and $Pr(y_D|H_1)$ are essentially probabilities of observing the data, after computing the Bayes factor for each data point, these individual Bayes factors can be multiplied to compute the overall Bayes factor under the assumption that the observations are independent, as
\begin{equation}
B = \prod_{i=1}^N{B_i}
\end{equation}

If the number of data points $N$ is relatively large and most of $B_i$'s are larger than one, the product of individual Bayes factors may also be a large number. In such a case it is more convenient to express Bayes factor on a logarithmic scale as
\begin{equation}
\log B = \sum_{i=1}^N \log B_i
\end{equation}

\paragraph{Interpretation of Bayesian hypothesis testing results}
If the Bayes factor calculated is greater than 1, it is indicated that the data favors the null hypothesis; if the Bayes factor is less than 1, it is indicated that the data favors the alternative hypothesis. In addition, Jef{}freys~\cite{Jeffreys1983} gave a heuristic interpretation of Bayes factor in terms of the level of support that the hypotheses obtain from data. The threshold value of Bayes factor $B_{th}$ can be related to the so-called Bayes risk in detection theory~\cite{Kay1998,Jiang2007}, which is the sum of costs due to dif{}ferent decision scenarios - failing to reject the true/wrong hypothesis and rejecting the true/wrong hypothesis. It has been shown that appropriate selection of $B_{th}$ can help reduce the Bayes risk~\cite{Kay1998}. If one assumes that the cost of making correct decisions (failing to reject the true hypothesis or rejecting the wrong hypothesis) is zero, the costs of type I and type II error are the same, and the prior probabilities of the null and alternative hypothesis being true are equal, then the resulting $B_{th}=1$~\cite{Jiang2007}. However, It should be noted that as a part of the decision making process, the choice of thresholds for Bayes factor inevitably contains subjective elements.

Before collecting validation data, there may be no evidence to support or reject the model. In that case, it may be reasonable to assume that the prior probabilities of the null hypothesis and alternative hypothesis are equal ($= 0.5$). In that case, a simple expression of the posterior probability of the null hypothesis can be derived in terms of the Bayes factor~\cite{Rebba2006}, which is a convenient metric to assess the conf{}idence in the model prediction:
\begin{eqnarray}\label{eq:BHTposteriorH0}
Pr(H_0|y_D) &=& \frac{Pr(y_D|H_0)Pr(H_0)}{Pr(y_D|H_0)Pr(H_0)+Pr(y_D|H_1)Pr(H_1)}\nonumber\\
&=& \frac{Pr(y_D|H_0)}{Pr(y_D|H_0)+Pr(y_D|H_1)}\\
&=& \frac{B}{1+B}\nonumber
\end{eqnarray}

An advantage of Bayesian hypothesis testing is that the posterior probabilities of $H_0$ and $H_1$ obtained from the validation exercise can both be used through a Bayesian model-averaging approach~\cite{Hoeting1999,Zhang2000,Zhang2003a} to ref{}lect the ef{}fect of the model validation result on the uncertainty in model output, as shown in Eq.~\ref{eq:BMA}
\begin{equation}\label{eq:BMA}
\bar{\pi}(y) = \pi_0(y) Pr(H_0|y_D) + \pi_1(y) Pr(H_1|y_D)
\end{equation}
where $\bar{\pi}(y)$ is the predicted PDF of $Y$ combining the PDFs of $Y$ under the null and alternative hypothese. Therefore, instead of completely accepting a single model, one can include the risk of using this model in further calculations. This helps to avoid both Type I and Type II errors, i.e., accepting a wrong model or rejecting a correct model.
\subsection{Relationship between $p$-value and Bayes factor}\label{section:relationBFandpvalue}
Although the $p$-value in classical hypothesis testing and the Bayes factor $B$ are based on dif{}ferent philosophical assumptions and formulated dif{}ferently, it has been shown that these two metrics can be mathematically related for some special cases~\cite{Rouder2009}. In the discussion below, the Bayes factor based on the hypothesis of probability density functions for a fully characterized experiment is found related to the $p$-value in $t$-test and $z$-test, if the model prediction $Y_m$ is a normal random variable with mean $\mu_m$ and standard deviation $\sigma_m$.

Starting from the formula of Bayes factor in Eq.~\ref{eq:BFvalidation}, since we assume that the PDF of the quantity to be predicted $Y$ under the alternative hypothesis $H_1$ is uniform, the integration term in the denominator is not af{}fected by the target model and thus can be treated as a constant $1/C$. Based on the null hypothesis $H_0$, the quantity to be predicted $Y \sim N(\mu_m,\sigma_m^2)$. Recall the relationship $Y_D = Y + \varepsilon_D$, and $\varepsilon_D \sim N(0,\sigma_D^2)$, we know that $Y_D \sim N(\mu_m, \sigma_m^2+\sigma_D^2)$. Thus the numerator of Eq.~\ref{eq:BFvalidation} can be calculated as
\begin{equation}
\int \pi(y_D|y)\pi_0(y|\boldsymbol{x})dy = \frac{1}{\sqrt{\sigma_m^2+\sigma_D^2}} \phi (\frac{y_D-\mu_m}{\sqrt{\sigma_m^2+\sigma_D^2}})
\end{equation}
where $\phi(*)$ is the PDF of the standard norm random variable.

If the variance of measurement noise is negligible compared to the variance of $Y_m$, i.e., $\sigma_D^2 \ll \sigma_m^2$, we have $\sigma_m^2+\sigma_D^2 \approx \sigma_m^2$. Also note that for a single data point $\bar{Y}_D=y_D$. Therefore Eq.~\ref{eq:BFvalidation} becomes
\begin{equation}\label{eq:BFsimp2}
B = \frac{C}{\sigma_m}*\phi(\frac{\bar{Y}_D-\mu_m}{\sigma_m})
\end{equation}

Based on Eqs.~\ref{eq:tstatistic} and~\ref{eq:zstatistic}, we have
\begin{eqnarray}\label{eq:diffmuANDtz}
\bar{Y}_D-\mu_m &=&
\begin{cases} 
t * S_D /\sqrt{n} & \text{, for $t$-test} \\
z * \sigma_{Y_D}/\sqrt{n} & \text{, for $z$-test}
\end{cases}
\end{eqnarray}

Substituting Eq.~\ref{eq:diffmuANDtz} into Eq.~\ref{eq:BFsimp2}, we obtain
\begin{eqnarray}\label{eq:BFandtz}
B &=&
\begin{cases}
C/\sigma_m*\phi[(t*S_D)/(\sigma_m\sqrt{n})] & \text{, for $t$-test} \\
C/\sigma_m*\phi[(z*\sigma_{Y_D})/(\sigma_m \sqrt{n})] & \text{, for $z$-test}
\end{cases}
\end{eqnarray}
where $\phi$ is the probability density function of a standard normal variable.

From Eq.~\ref{eq:BFandtz}, we can see that the Bayes factor can be related to either the $z$ statistic or the $t$ statistic, and hence it can be related to the $p$-value in both $z$-test and $t$-test. Combining Eqs.~\ref{eq:pvalueInztest} and~\ref{eq:BFandtz}, we obtain the relation between Bayes factor and the $p$-value in the $z$-test as
\begin{equation}\label{eq:BFandpvalueInztest}
B = \frac{C}{\sigma_m}*\phi[\Phi^{-1}(\frac{p}{2}) \frac{\sigma_{Y_D}}{\sigma_m \sqrt{n}}]
\end{equation}
where $\Phi^{-1}$ is the inverse standard normal CDF. Similarly, the relation between Bayes factor and the $p$-value in the $t$-test can be obtained by combining Eqs.~\ref{eq:pvalueInttest} and~\ref{eq:BFandtz} as
\begin{equation}\label{eq:BFandpvalueInttest}
B = \frac{C}{\sigma_m}*\phi\{[S_D*F_{T,n-1}^{-1}(\frac{p}{2})]/(\sigma_m\sqrt{n})\}
\end{equation}
where $F_{T,n-1}^{-1}$ is the inverse CDF of a $t$-distribution with $(n-1)$ degrees of freedom.

If the chosen signif{}icance level in $z$-test or $t$-test is $\alpha$, the corresponding threshold Bayes factor $B_{th}$ can be calculated using Eq.~\ref{eq:BFandpvalueInztest} or~\ref{eq:BFandpvalueInttest} by letting $p=\alpha$. In that case, the $z$-test/$t$-test with signif{}icance level $\alpha$ and Bayesian hypothesis testing with the corresponding threshold value $B_{th}$ will both give the same model validation result.
\section{Non-hypothesis testing-based methods}\label{section:nonHT}
Besides the binary hypothesis testing methods discussed above, various other validation metrics have been developed to quantify the agreement between models and experimental data from other perspectives, such as the Mahalanobis distance for models with multivariate output~\cite{Srivastava2002}, the weighted validation data-based metric~\cite{Hills2003}, the Kullback-Leibler divergence in the area of signal processing~\cite{Seghouane2005} and for the design of validation experiments~\cite{Jiang2006}, the probability bound-based metric~\cite{Urbina2003}, the conf{}idence interval-based metric~\cite{Oberkampf2006}, the reliability-based metric~\cite{Rebba2008}, and the area metric~\cite{Ferson2009,Ferson2008}. The weighted validation data-based metric introduced by Hills and Leslie~\cite{Hills2003} is designed for the case when the importance of dif{}ferent validation experiments is dif{}ferent. The conf{}idence interval-based validation method proposed by Oberkampf et al.~\cite{Oberkampf2006} computes the conf{}idence interval of error, which is def{}ined as the dif{}ference between the model mean prediction and the true mean of the variable to be predicted. An average absolute error metric and an average absolute conf{}idence indicator are also computed. However, it is not clear how to apply this method to validation of a multivariate model with data from discrete test combinations, as the method requires integration over the space of test inputs. Therefore, only the reliability-based metric and the area metric are discussed in this paper.
\subsection{Reliability-based metric}\label{section:reliabilityMetric}
The reliability metric $r$ proposed by Rebba and Mahadevan~\cite{Rebba2008} is a direct measure of model prediction quality and is relatively easy to compute. It is def{}ined as the probability of the dif{}ference $d$ between model prediction and observed data being less than a given quantity $\epsilon$
\begin{equation}
r = Pr(-\epsilon<d<\epsilon) \qquad d=Y_D-Y_m
\end{equation}

As mentioned in Section~\ref{section:validationMethods}, experimental observation is random due to measurement uncertainty and model output may be uncertain in the Bayesian framework. Therefore, as the difference between two random variables, in the Bayesian framework we interpret $d$ as a random variable. The probability distribution of $d$ can be obtained from the probability distributions of $Y_D$ and $Y_m$. For instance, if the model prediction $Y_m \sim N(\mu_m,\sigma_m^2)$, and the corresponding observation $Y_D \sim N(\mu,\sigma_{Y_D}^2)$ (see discussion in Section~\ref{section:CHT}), the dif{}ference $d \sim N(\mu-\mu_m,\sigma^2+\sigma_D^2+\sigma_m^2)$. For the sake of simplicity, let $\sigma_d=\sqrt{\sigma^2+\sigma_D^2+\sigma_m^2}$. In this case, the reliability-based metric $r$ can be derived as
\begin{equation}\label{eq:rmformula}
r = \Phi[\frac{\epsilon-(\mu-\mu_m)}{\sigma_d}]-\Phi[\frac{-\epsilon-(\mu-\mu_m)}{\sigma_d}]
\end{equation}

In this paper, experimental data are considered as the samples of the random variable $Y_D$. Therefore, if the number of experimental data is relatively large, e.g., $n>30$, the sample variance $S_D^2$ can be assumed as a good estimator of $\sigma_{Y_D}^2$ (the true variance of $Y_D$), which is needed to compute the reliability metric. If $n$ is small and no prior information on $\sigma$ is available, we can assume that $\sigma=\sigma_m$, which is the same assumption used in $z$-test. By assuming further that the mean of validation data $\bar{Y}_D$ is equal to $\mu$, Eq.~\ref{eq:rmformula} can be re-written as
\begin{eqnarray}\label{eq:rmformula2}
r &=& \Phi[\frac{\epsilon-(\bar{Y}_D-\mu_m)}{\sigma_d}]-\Phi[\frac{-\epsilon-(\bar{Y}_D-\mu_m)}{\sigma_d}]
\end{eqnarray}

By substituting Eq.~\ref{eq:diffmuANDtz} into Eq.~\ref{eq:rmformula2}, the relation between the reliability-based metric $r$ and the test statistic in the $t$-test or $z$-test is obtained as
\begin{eqnarray}\label{eq:rANDtz}
r &=&
\begin{cases}
\Phi[(\epsilon-t*S_D/\sqrt{n})/\sigma_d]+\Phi[(\epsilon+t*S_D/\sqrt{n})/\sigma_d]-1 & \text{, for $t$-test} \\
\Phi[(\epsilon-z*\sigma_{Y_D}/\sqrt{n})/\sigma_d]+\Phi[(\epsilon+z*\sigma_{Y_D}/\sqrt{n})/\sigma_d]-1 & \text{, for $z$-test}
\end{cases}
\end{eqnarray}

By combining Eqs.~\ref{eq:pvalueInttest},~\ref{eq:pvalueInztest} and~\ref{eq:rANDtz}, the reliability-based metric can be further related to the $p$-value in the $t$-test or $z$-test as
\begin{eqnarray}\label{eq:rmAndp}
r &=& 
\begin{cases}
\Phi[(\epsilon-F_{T,n-1}^{-1}(p/2)*S_D/\sqrt{n})/\sigma_d]+ \\
\qquad \Phi[(\epsilon+F_{T,n-1}^{-1}(p/2)*S_D/\sqrt{n})/\sigma_d]-1 & \text{, for $t$-test} \\
\Phi[(\epsilon-\Phi^{-1}(p/2)*\sigma_{Y_D}/\sqrt{n})/\sigma_d]+ \\
\qquad \Phi[(\epsilon+\Phi^{-1}(p/2)*\sigma_{Y_D}/\sqrt{n})/\sigma_d]-1 & \text{, for $z$-test}
\end{cases}
\end{eqnarray}

If one chooses to test models based on a threshold reliability value $r_{th}$ calculated by letting $p=\alpha$ in Eq.~\ref{eq:rmAndp} above, the result of model validation will be the same as that in the $t$-test or $z$-test with signif{}icance level $\alpha$.

Note that the threshold $r_{th}$ used in the reliability-based method represents the minimum probability of the dif{}ference $d$ falling within an interval [$-\epsilon,\epsilon$], and the decision of accepting/rejecting a model can be made based on the decision maker's acceptable level of model reliability. 

Since the reliability-based metric is the probability of $d$ being within a given interval, it can also ref{}lect the existence of directional bias by modifying the intervals. Similar to the Bayesian interval hypothesis testing, we can take two dif{}ferent intervals $[0,\epsilon]$ and $[-\epsilon,0]$, and calculate the corresponding values of metric $r^1$ and $r^2$ as:
\begin{eqnarray}\label{eq:rmformula1}
r^1 &=& \Phi[\frac{\epsilon-(\mu-\mu_m)}{\sigma_d}]-\Phi[\frac{-(\mu-\mu_m)}{\sigma_d}] \nonumber\\
r^2 &=& \Phi[\frac{-(\mu-\mu_m)}{\sigma_d}]-\Phi[\frac{-\epsilon-(\mu-\mu_m)}{\sigma_d}]
\end{eqnarray}

By comparing the values of $r^1$ and $r^2$ with the threshold $r_{th}/2$ (half of the original threshold value because the width of intervals considered is half of the original one), the model will be failed if either $r^1$ or $r^2$ is less than $r_{th}/2$.

Note that for the case that the quantity to be predicted $Y$ is deterministic, $\sigma$ becomes zero, and for the case that the model prediction $Y_m$ is deterministic, $\sigma_m$ becomes zero.

\subsection{Area metric-based method}\label{section:areaMetric}
The area metric proposed by Ferson et al.~\cite{Ferson2009,Ferson2008} is attractive due to its capability to incorporate fully characterized experiments using the so-called "u-pooling" procedure, and thus to validate models with sparse data on multiple experimental combinations~\cite{Liu2011a}. For a single experimental combination with input $\boldsymbol{x_i}$, suppose $F_{\boldsymbol{x_i}}^m$ is the corresponding cumulative probability function (CDF) of model output $Y_m$ and $y_{Di}$ is the observed data, one can compute a $u$-value for this experimental combination as
\begin{equation}\label{eq:upooling}
u_i = F_{\boldsymbol{x_i}}^m(y_{Di})
\end{equation}

Based on the probability integral transform theory in statistics~\cite{Angus1994}, if the observed data $y_{Di}$ is a random sample from the probability distribution of model output, the computed $u_i$ will be a random sample from the standard uniform distribution, and thus the empirical CDF of all the $u_i$'s ($i = 1, 2, ..., N$) should match the CDF of the standard uniform random variable. The dif{}ference between these two CDF curves is a measure of the disparity between model predictions and experimental observations. Hence, a model validation metric can be developed as~\cite{Ferson2008}
\begin{equation}\label{eq:areametric}
d(F_u,S_u) = \int_0^1 |F_u-S_u|du
\end{equation}
where $F_u$ is the empirical CDF of all the $u_i$'s and $S_u$ is the CDF of the standard uniform random variable. If the value of $d(F_u,S_u)$ is small/large, the model predictions are correspondingly close/not close to experimental observations.

If the model prediction $Y_m$ is deterministic, the CDF function $F_{\boldsymbol{x_i}}^m$ in Eq.~\ref{eq:upooling} cannot be constructed, and hence the area metric-based method is not applicable to testing a computational model with deterministic output.

The area metric can also ref{}lect the existence of directional bias, i.e., the experimental observations are consistently below or above the corresponding mean predictions of numerical model. This is due to the use of CDF of model output in Eq.~\ref{eq:upooling}. For example, if the model outputs at dif{}ferent test combinations are Gaussian random variables, the $F_{\boldsymbol{xi}}^m$'s in Eq.~\ref{eq:upooling} become Gaussian CDFs. Hence, the values of $F_{\boldsymbol{xi}}^m(y_{Di})$ will all be less than 0.5 if $y_{Di}$'s are smaller than the mean of the corresponding Gaussian variables. Therefore, instead of being uniformly distributed between [0,1], $u_i$'s are distributed between [0,0.5], causing a large area between the empirical CDF of $u_i$ and the standard uniform CDF.

Compared to the hypothesis testing methods and the reliability-based method, the area metric-based method lacks a direct interpretation of model acceptance threshold, i.e., it is not clear how to set up an appropriate threshold to decide when one should reject/accept a model. This is a disadvantage for the area metric-based approach.
\section{Numerical example}\label{section:numericalExample}
In this section, the aforementioned model validation methods are illustrated via an application example on damping prediction for MEMS switches. The quantity to be predicted, the damping coef{}f{}icient, is considered as a random variable due to the lack of understanding in physical modeling, in other words, the epistemic uncertainty of damping coef{}f{}icient is represented by a subjective probability distribution following the Bayesian way of thinking; the corresponding computational model is also stochastic as will be shown in Section~\ref{section:UQinDamping}. The validation data are obtained from fully characterized experiments, and it is found that the directional bias def{}ined in Section~\ref{section:intro} exists between model prediction and validation data.

\subsection{Damping model and experimental data}
Despite the superior performance provided in terms of signal loss and isolation compared with silicon devices~\cite{Rebeiz2003}, the use of RF MEMS switches in applications requiring high reliability is hindered by signif{}icant variations in device lifetime~\cite{Guo2010}. Rigorous quantif{}ication of the uncertainty sources contributing to the observed life variations is necessary in order to achieve the design of reliable devices. Within the framework of uncertainty quantif{}ication in the modeling of RF MEMS switches, the validation of squeeze-f{}ilm damping model emerges as a crucial issue due to two factors: (1) damping strongly af{}fects the dynamic behavior of the MEMS switch and therefore its lifetime~\cite{Snow2010}; (2) it is dif{}f{}icult to accurately model micro-scale f{}luid damping and available models are applicable to limited regimes~\cite{Bidkar2009}.
\subsubsection{Uncertainty quantif{}ication in micro-scale squeeze-f{}ilm damping prediction}\label{section:UQinDamping}
For the purpose of illustration, this study considers damping prediction using the Navier-Stokes slip jump model~\cite{Gad-el-Hak2005}. Two major sources of uncertainty have been shown to af{}fect the prediction of gas damping~\cite{Guo2010}. The f{}irst one is epistemic uncertainty related to the lack of understanding of fundamental failure modes and related physical models. The second one is aleatory uncertainty in model parameters and inputs due to variability in either the fabrication process or in the operating environment. Uncertainty quantif{}ication approaches usually require large numbers of deterministic numerical simulations. In order to reduce the computational cost, a generalized polynomial chaos (gPC) surrogate model~\cite{Xiu2002} is constructed and trained using solutions of the Navier-Stokes (N-S) equation for a few input points, thus avoiding repetitively solving the N-S equation. Note that several other surrogate modeling techniques are also available, including Kriging or Gaussian Process (GP) interpolation~\cite{Rasmussen2006}, support vector machine (SVM)~\cite{Vapnik1999}, relevance vector machine~\cite{Tipping2001}, etc. The gPC model is used for the purpose of illustration. This model approximates the target stochastic function using orthogonal polynomials in terms of the random inputs~\cite{Guo2010}. A $P^{th}$ order gPC model $y_m(\boldsymbol{x})$ that approximates a random function $y(\boldsymbol{x})$ can be written as
\begin{equation}\label{eq:gPCmodel}
y(\boldsymbol{x}) \approx y_m(\boldsymbol{x}) = \sum_{i=1}^M a_i\phi_i(\boldsymbol{x}) + \varepsilon_m\qquad M = \begin{pmatrix}
n_x + P \\
n_x
\end{pmatrix}
\end{equation}
where $\phi_i$'s are the orthonormal polynomial bases such as Legendre polynomials, Hermite polynomials, and Wiener-Askey polynomials; $n_x$ is the dimension of input $\boldsymbol{x}$ and $P$ is the order of the polynomial; $\varepsilon_m$ is the error of the gPC model; $a_i$'s are coef{}f{}icients and can be obtained as
\begin{equation}\label{eq:gPCcoeff}
a_i = \frac{\int y_m(\boldsymbol{x})\phi_i(\boldsymbol{x})d\boldsymbol{x}}{\int \phi_i^2(\boldsymbol{x})d\boldsymbol{x}} = \frac{1}{h_i}\sum_{j=1}^N w_jy(\boldsymbol{x_j})\phi_i(\boldsymbol{x_j})
\end{equation}
where $h_i=\int \phi_i^2(\boldsymbol{x})d\boldsymbol{x}$ is constant for a given polynomial basis $\phi_i(\boldsymbol{x})$, and $\{\boldsymbol{x_j},w_j\}_{j=1}^N$ is a set of nodes and weights of the quadrature rule for numerical integration.

Based on the calculated damping coef{}f{}icient values $y(\boldsymbol{x_j})$ at the quadrature nodes $\boldsymbol{x_j}$ by solving the Navier-Stokes Slip Jump model, the gPC model $y_m(\boldsymbol{x})$ can be constructed using Eqs.~\ref{eq:gPCmodel} and~\ref{eq:gPCcoeff}. For any given input $\boldsymbol{x_k}$, $\mu_m(\boldsymbol{x_k})=\sum_{i=1}^M a_i\phi_i(\boldsymbol{x_k})$ is deterministic, while the residual term $\varepsilon_m$ is random. Under the Gauss-Markov assumption, $\varepsilon_m$ asymptotically follows a Gaussian distribution with zero mean, and the variance can be estimated as~\cite{Liang2011,Seber2003}
\begin{equation}\label{eq:gPCresidual}
\sigma_m^2 = \sigma^2[1+\boldsymbol{\phi}^T(\boldsymbol{x_k})(\boldsymbol{\Phi}^T\boldsymbol{\Phi})^{-1}\boldsymbol{\phi}(\boldsymbol{x_k})]
\end{equation}
where $\sigma_m^2$ is a function of model input $\boldsymbol{x_k}$; the vector $\boldsymbol{\phi}(\boldsymbol{x_k})$ = [$\phi_1(\boldsymbol{x_k})$,$\phi_2(\boldsymbol{x_k})$,...,$\phi_M(\boldsymbol{x_k})]^T$; the matrix $\boldsymbol{\Phi}$ = $[\boldsymbol{\phi}(\boldsymbol{x_1})$,$\boldsymbol{\phi}(\boldsymbol{x_2})$,...,$\boldsymbol{\phi}(\boldsymbol{x_N})]^T$; and $\sigma^2 = 1/(N-M)\sum_{j=1}^N [\mu_m(\boldsymbol{x_j})-y(\boldsymbol{x_j})]^2$.

Therefore, for a given input $\boldsymbol{x_k}$, the prediction of damping coef{}f{}icient based on the gPC model is a random variable with Gaussian distribution $N(\mu_m(\boldsymbol{x_k}),\sigma_m(\boldsymbol{x_k}))$. The methods presented in Sections~\ref{section:binHT} and~\ref{section:nonHT} will be applied to the validation of this predicted distribution.

\begin{figure}[h!]
\begin{center}
\includegraphics[width=0.9\textwidth]{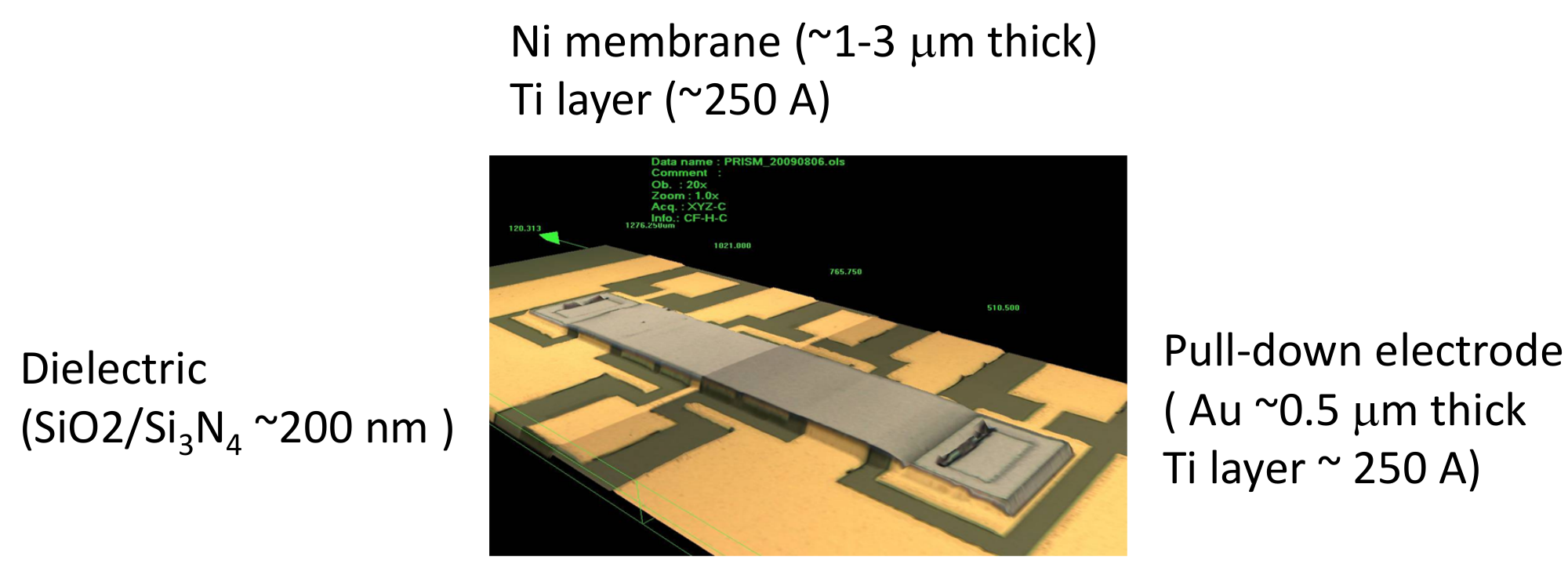}
\end{center}
\caption{Example RF MEMS switch (Courtesy: Purdue PRISM center)}
\label{fig:targetDevice}
\end{figure}

The example RF MEMS switch modeled as a membrane is shown in Fig.~\ref{fig:targetDevice}. To construct a gPC model for the damping coef{}f{}icient, the input parameters $\boldsymbol{x}$ need to be specif{}ied f{}irst. A probabilistic sensitivity analysis shows that the membrane thickness $t$, the gap height $g$, and the frequency $\omega$ are the major sources of variability in the damping coef{}f{}icient, and hence these three parameters are included in the gPC model, i.e., $\boldsymbol{x}=[t,g,\omega]$. Four dif{}ferent gas pressures - 18798.45 Pa, 28664.31 Pa, 43596.41 Pa, and 66661.19 Pa - are considered and correspondingly four gPC models are constructed. This example uses a third order gPC model with Legendre polynomial bases~\cite{Guo2010}.

It should be noted that the validity of the surrogate model does not guarantee the validity of the original model. We only have access to the surrogate model and validation experimental data; therefore in this example we are only assessing the validity of the surrogate model.

\subsubsection{Experimental data for validation}
In the experiment, seven devices with dif{}ferent geometric dimensions are considered. For each of the four pressures mentioned above, 5 repetitive tests are conducted on each of the seven devices, and hence 140 data points are obtained. Since the input parameters $[t,g,\omega]$ are recorded for each of the data points, these experiments are fully characterized. However, there is only one data point for each test combination due to the fact that each of the 140 input value combinations is dif{}ferent from others.

\begin{figure}[h!]
\begin{center}
\subfigure[]{
\includegraphics[width=0.46\textwidth]{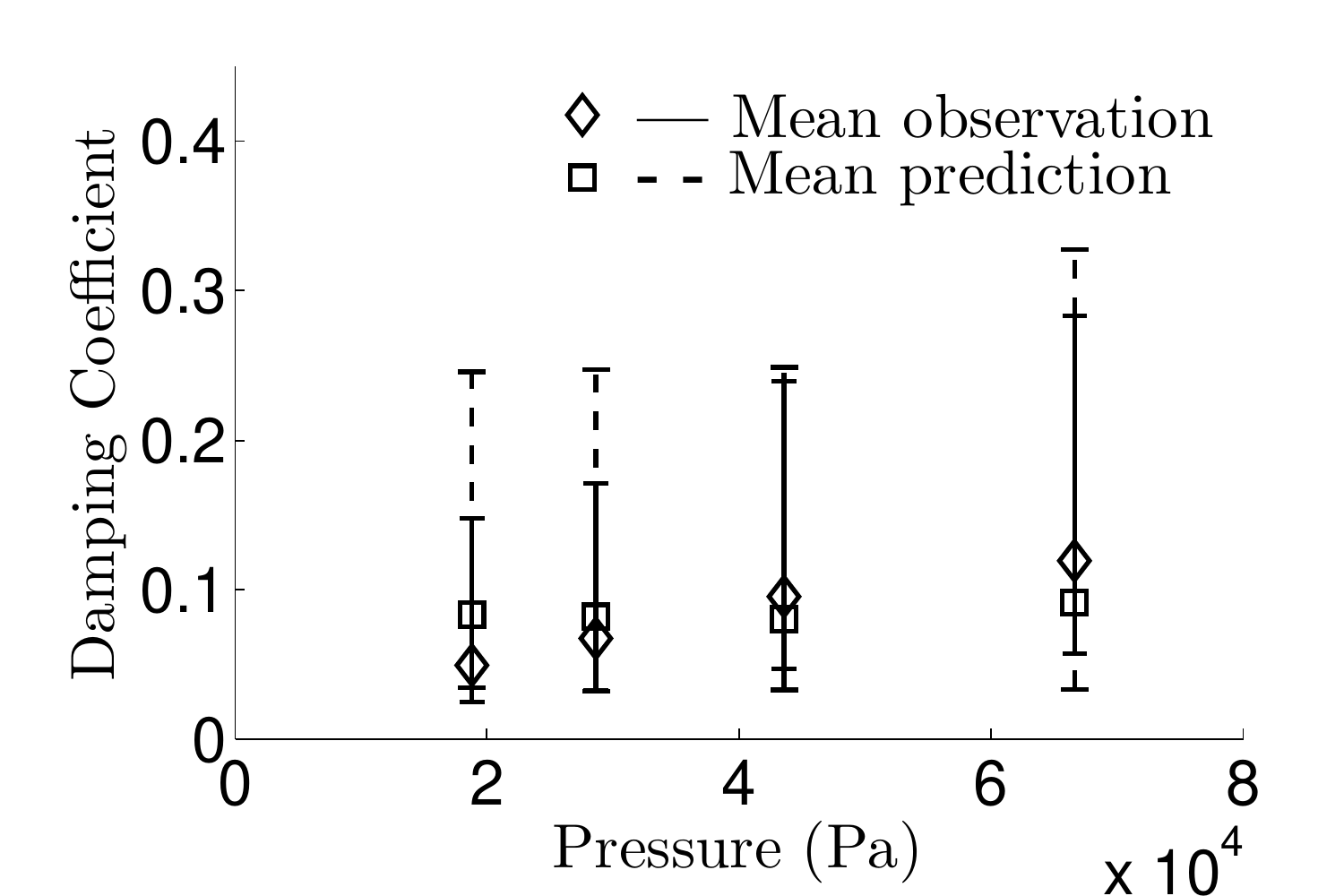}\label{fig:aggData}}\\
\subfigure[]{
\includegraphics[width=0.46\textwidth]{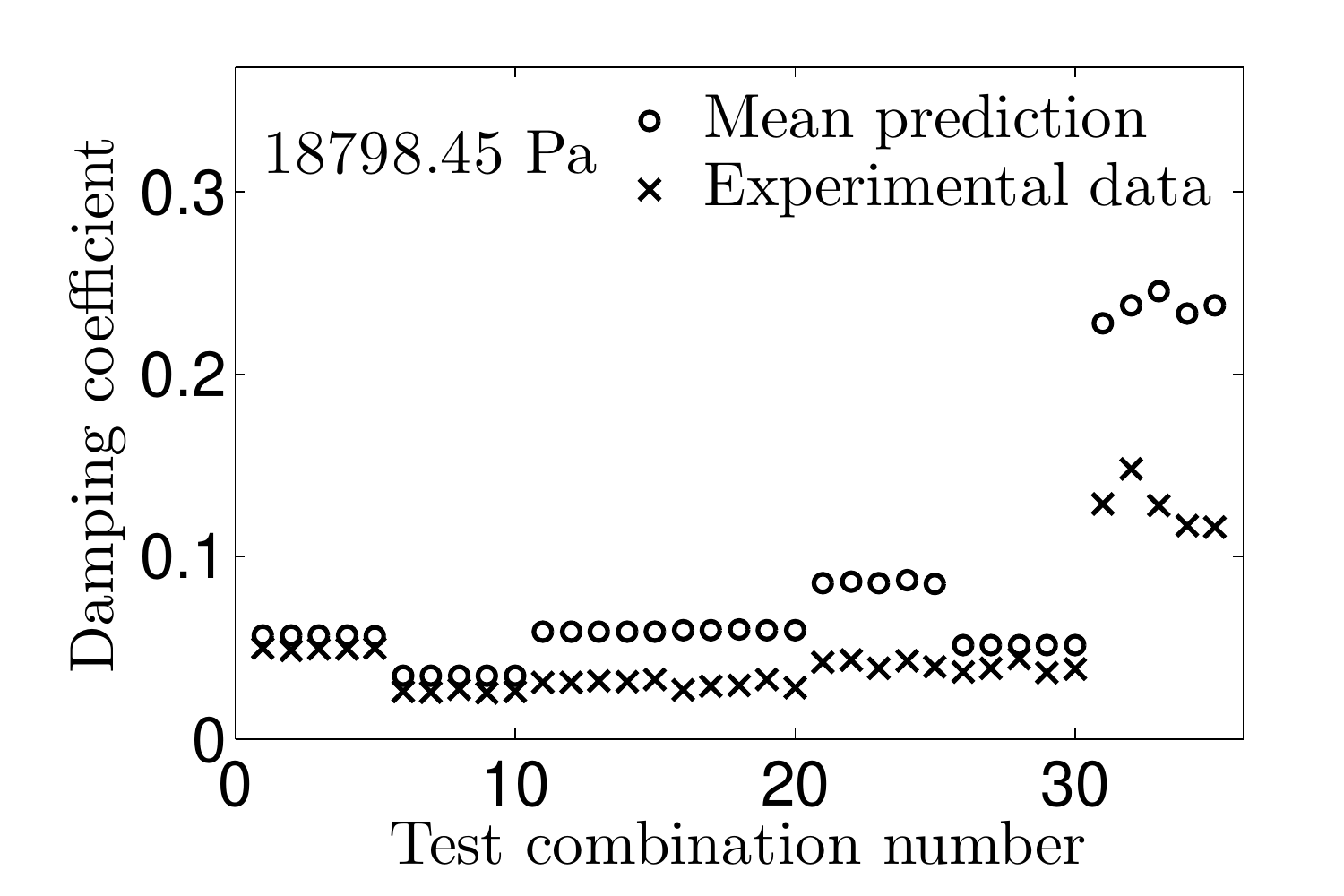}\label{fig:GraphP1}}
\subfigure[]{
\includegraphics[width=0.46\textwidth]{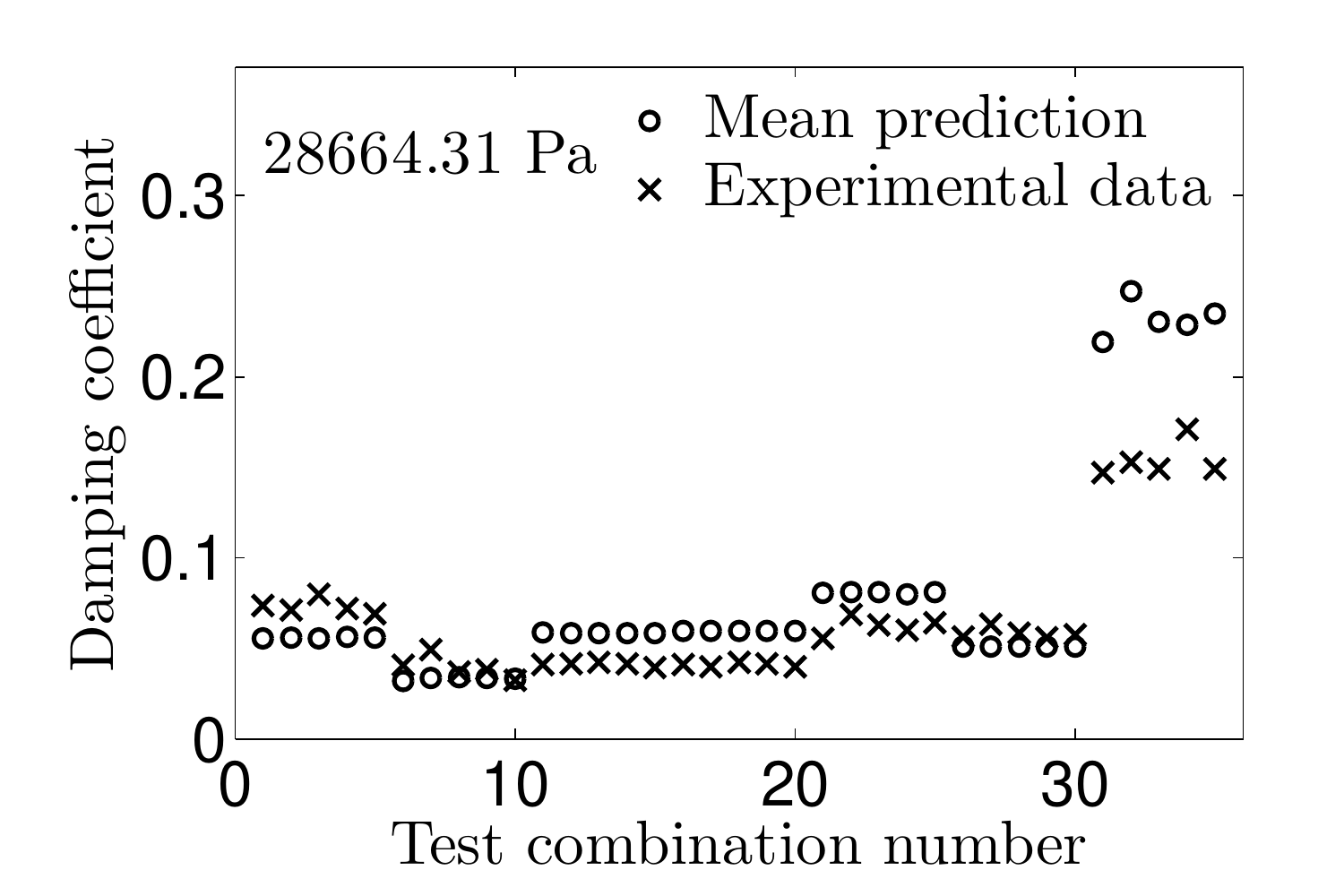}\label{fig:GraphP2}}\\
\subfigure[]{
\includegraphics[width=0.46\textwidth]{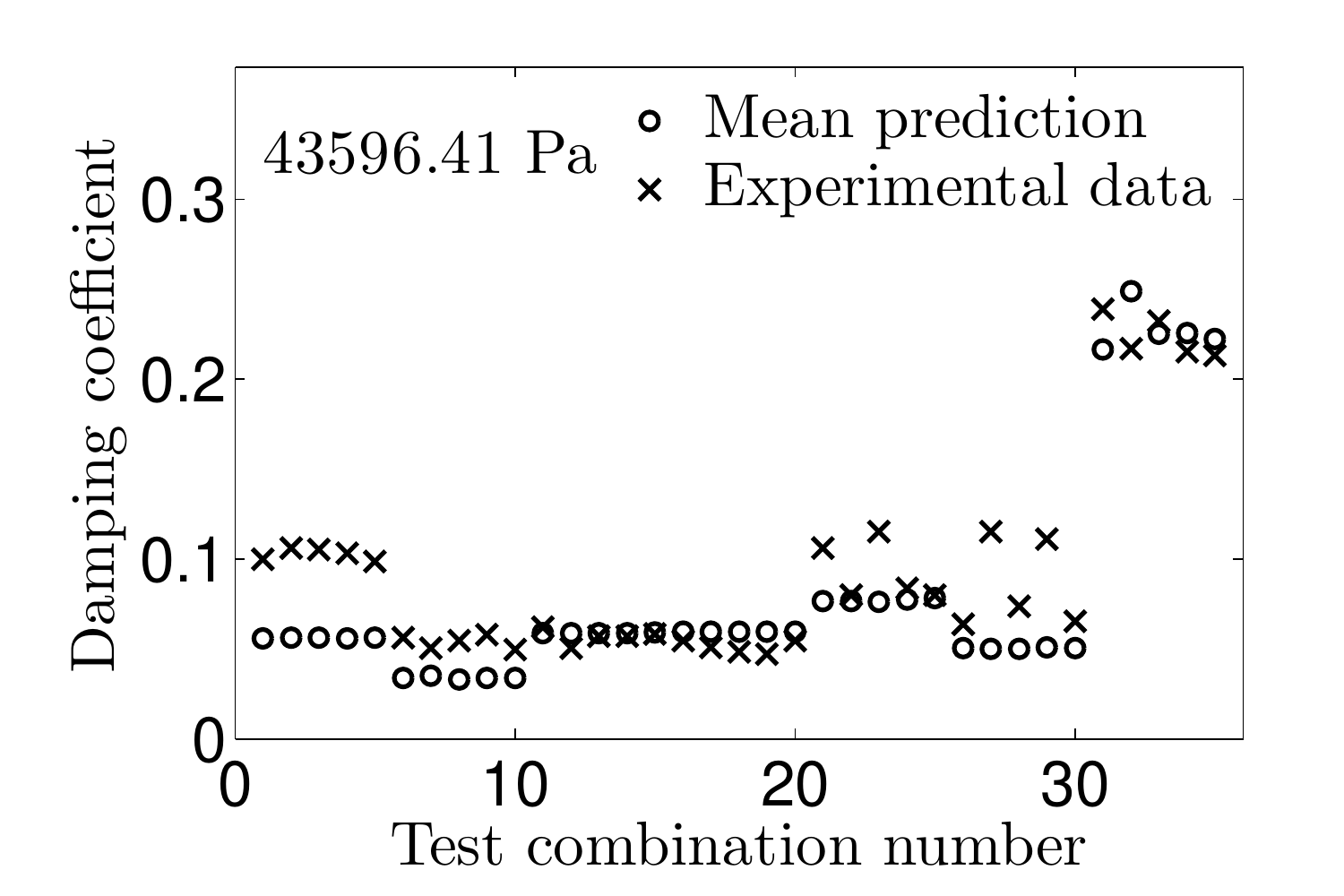}\label{fig:GraphP3}}
\subfigure[]{
\includegraphics[width=0.46\textwidth]{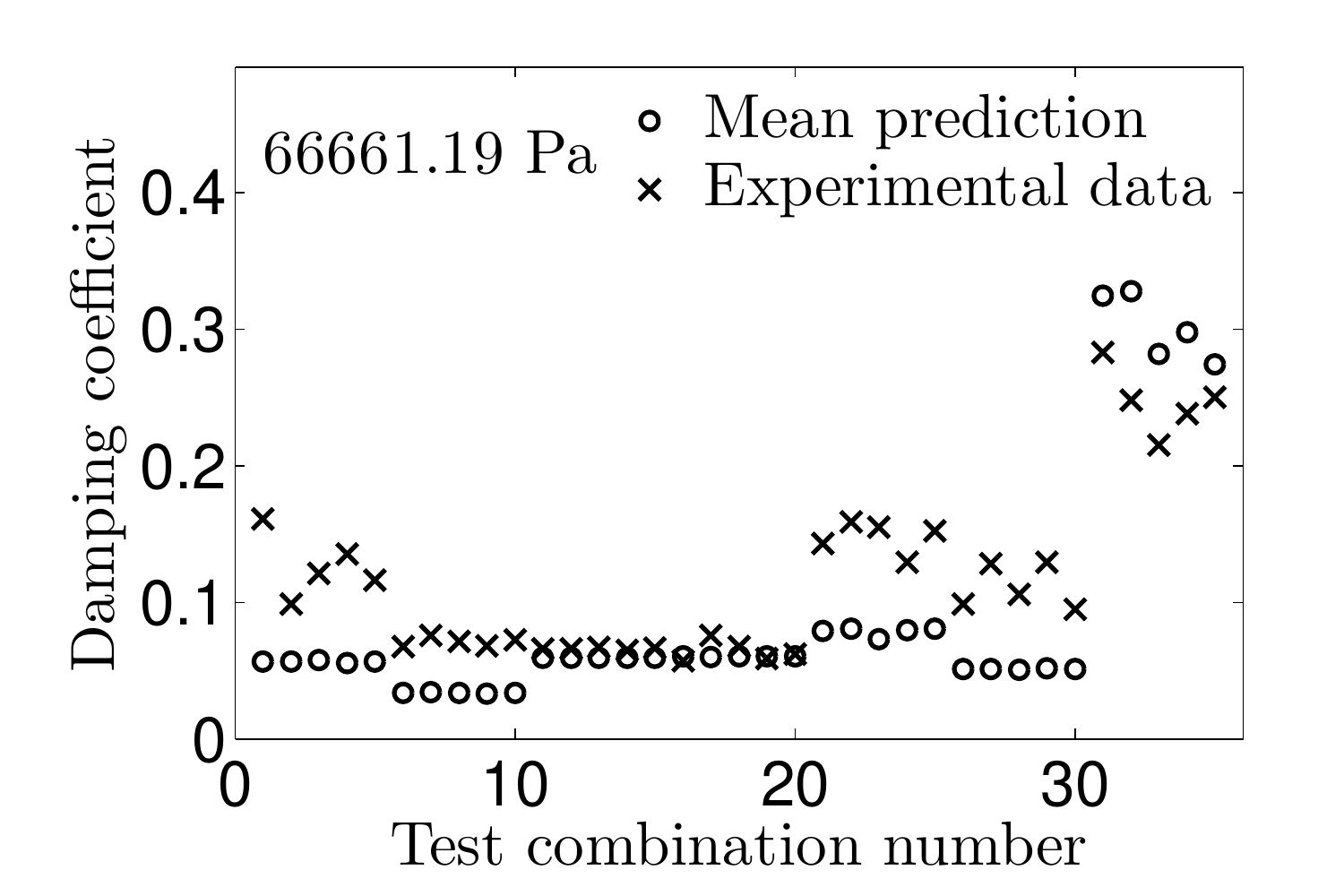}\label{fig:GraphP4}}
\end{center}
\caption{Graphical comparisons between gPC predictions and experimental data}
\end{figure}

Fig.~\ref{fig:aggData} shows a graphical comparison between the mean gPC model prediction and experimental data under the four dif{}ferent pressures by aggregating predictions and data with respect to the 35 test combinations for each pressure value. The top/bottom points are correspondingly the maximum/minimum value of model mean predictions and experimental data, and the square/diamond markers are the average values of predictions/data on the 35 test combinations. A more detailed graphical comparison showing mean prediction of the gPC model vs. experimental data on each of the individual test combinations is provided in Figs.~\ref{fig:GraphP1}-\subref{fig:GraphP4}.

From the graphical comparison, we can see that the gPC model performs better under the middle two values of pressure. Also note that there is a systematic bias between the gPC model and experimental observations at the low pressure value (18798.45 Pa), i.e., the mean predictions of the gPC model are always larger than the experimental data.
\subsection{Validation based on binary hypothesis testing}
\subsubsection{Classical hypothesis testing}\label{section:numExpCHT}
\begin{figure}[h!]
\begin{center}
\subfigure[]{
\includegraphics[width=0.46\textwidth]{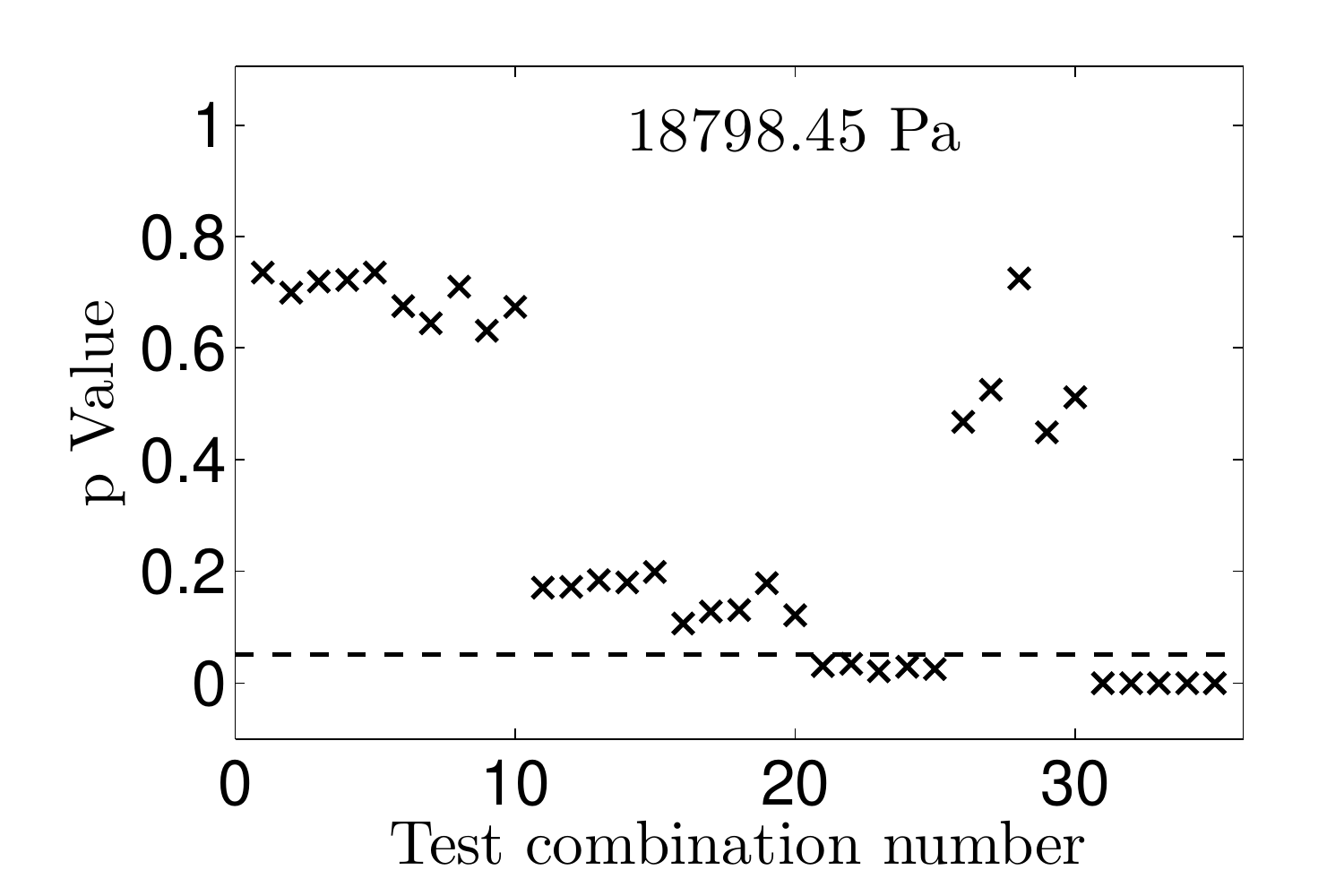}}
\subfigure[]{
\includegraphics[width=0.46\textwidth]{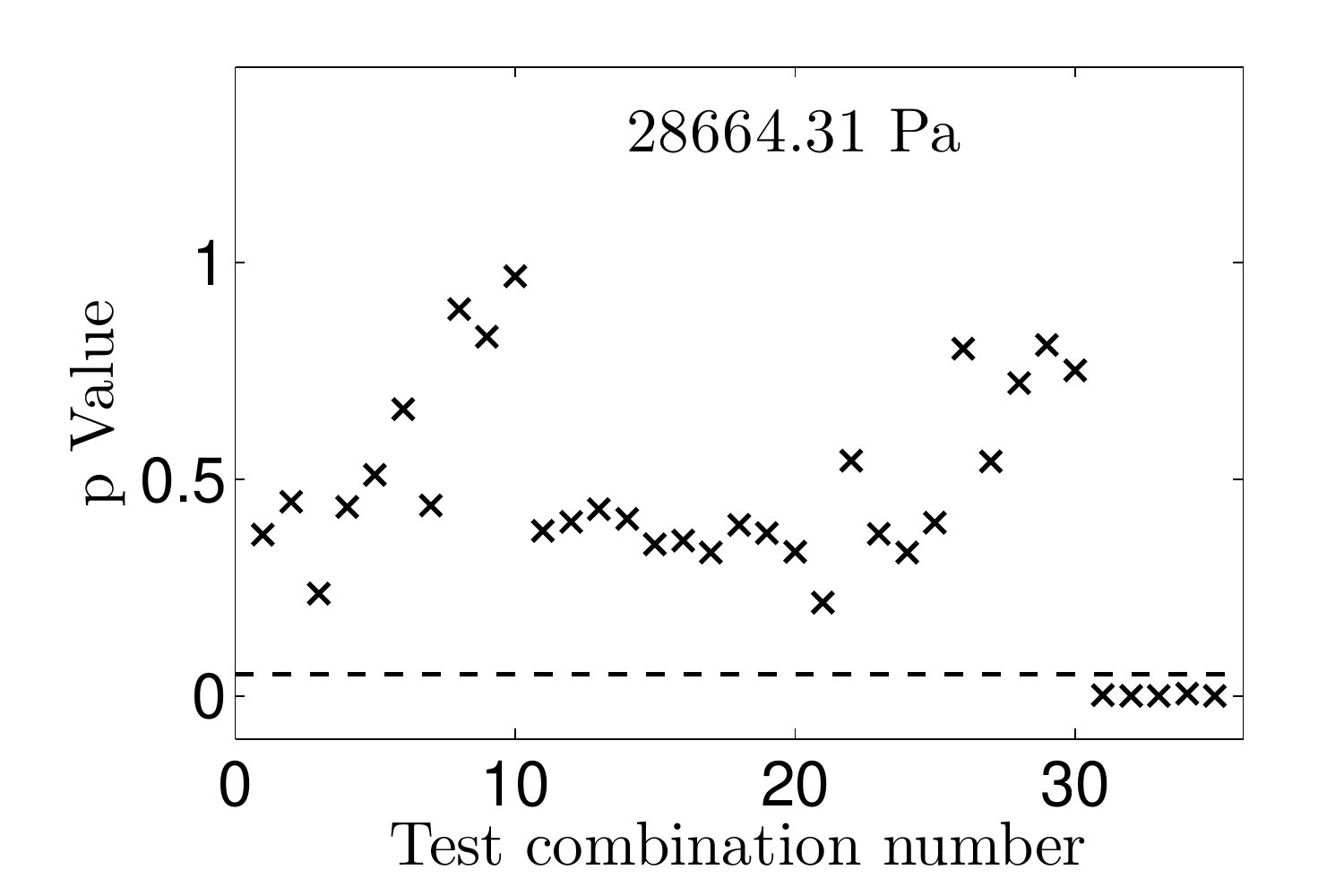}}\\
\subfigure[]{
\includegraphics[width=0.46\textwidth]{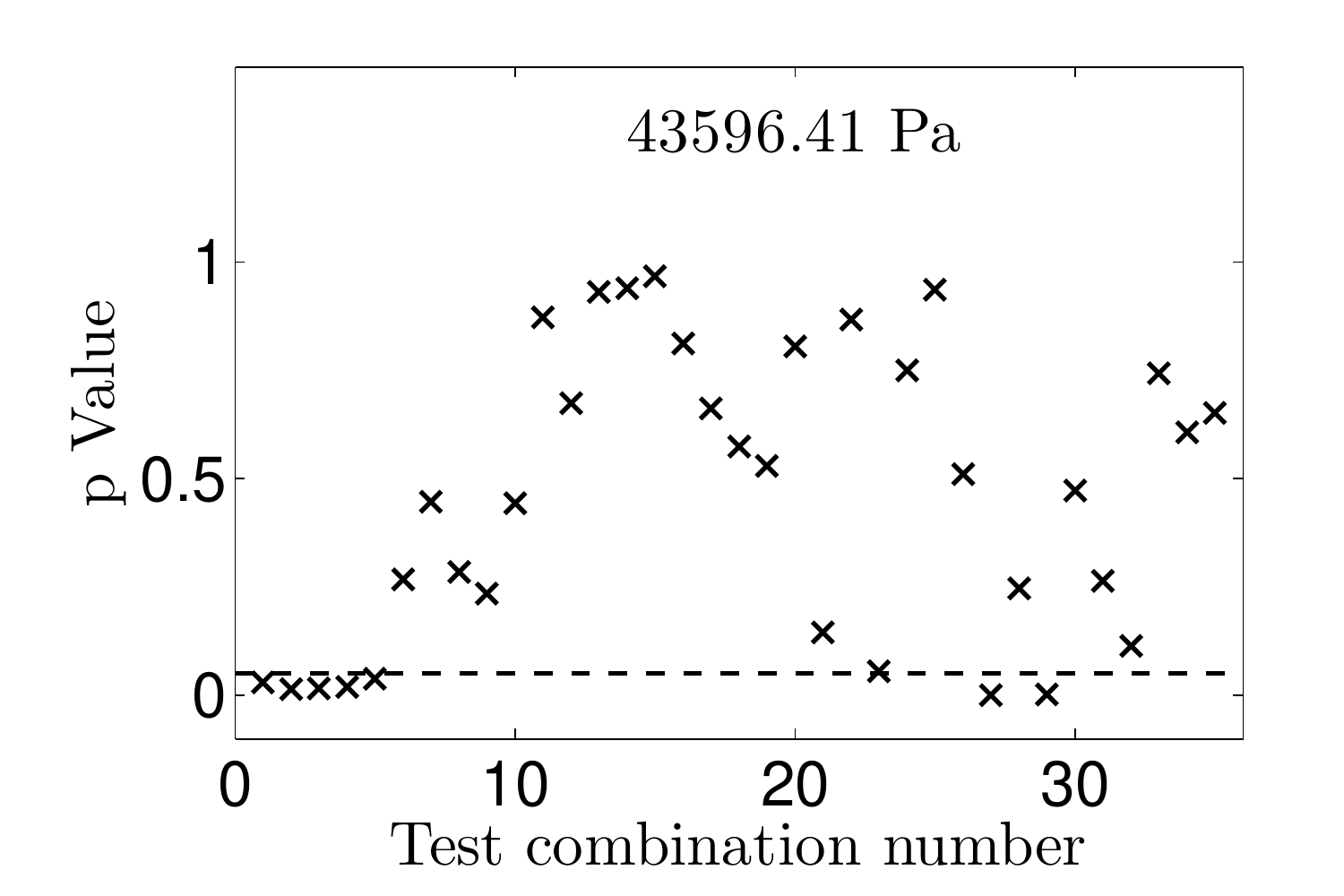}}
\subfigure[]{
\includegraphics[width=0.46\textwidth]{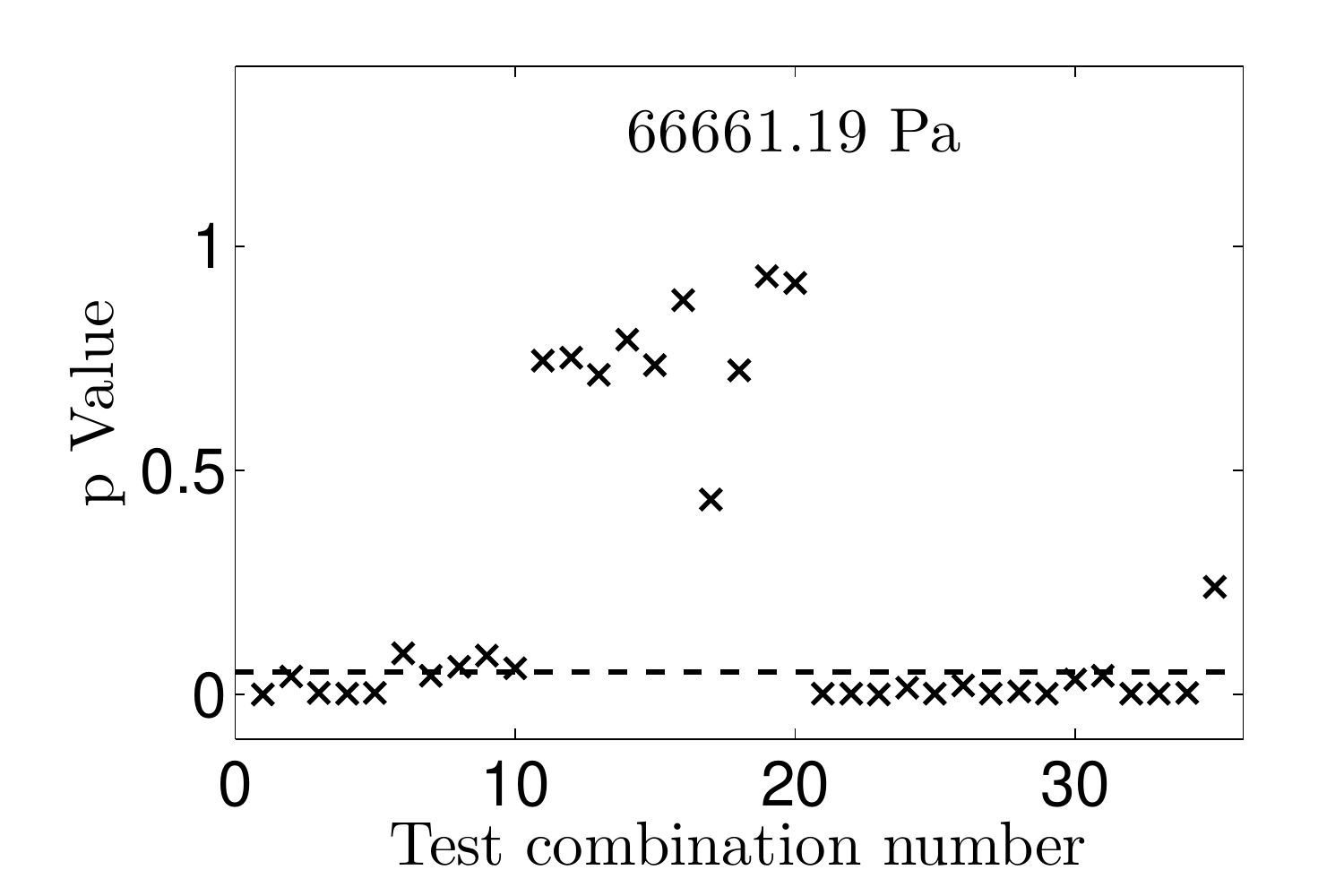}}
\end{center}
\caption{$p$-value of z-test}
\label{fig:pvalue}
\end{figure}

Because the sample size for each experimental combination is only 1, the $t$-test is not applicable and instead $z$-test is used. The $p$-values calculated using Eq.~\ref{eq:pvalueInztest} are shown in Fig.~\ref{fig:pvalue}. The dashed lines in Fig.~\ref{fig:pvalue} represent the signif{}icance level $\alpha = 0.05$. The model is considered to have failed at the experimental combinations where the corresponding $p$-values fall below the dashed line. Note that a more rigorous test will need to include the probability of making type II error ($\beta$). The individual numbers of failures of the four gPC models are shown in Table~\ref{table:gPcInztest}.

\begin{table}[h!]
\begin{center}
\caption{Performance of gPC models in $z$-test with $\alpha=0.05$}
\label{table:gPcInztest}
\begin{tabular}{lcccc}
\toprule
Pressure (Pa) & 18798.45 & 28664.31 & 43596.41 & 66661.19\\
\midrule
Number of failures & 10 & 5 & 7 & 20\\
Failure percentage & 28.6\% & 14.3\% & 20.0\% & 57.1\%\\
\bottomrule
\end{tabular}
\end{center}
\end{table}
\subsubsection{Bayesian hypothesis testing}\label{section:numExpBHT}
\begin{figure}[h!]
\begin{center}
\subfigure[]{
\includegraphics[width=0.46\textwidth]{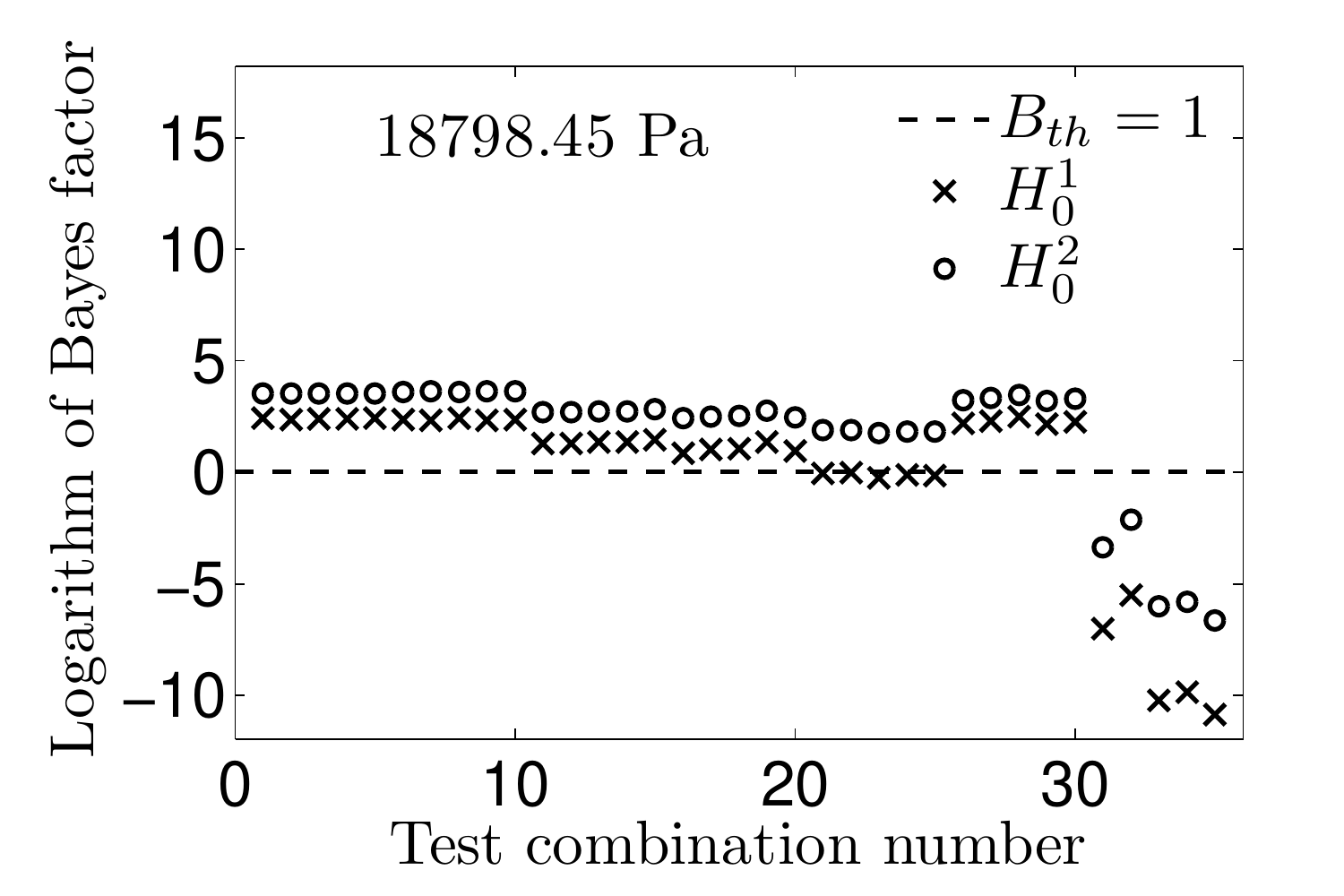}}
\subfigure[]{
\includegraphics[width=0.46\textwidth]{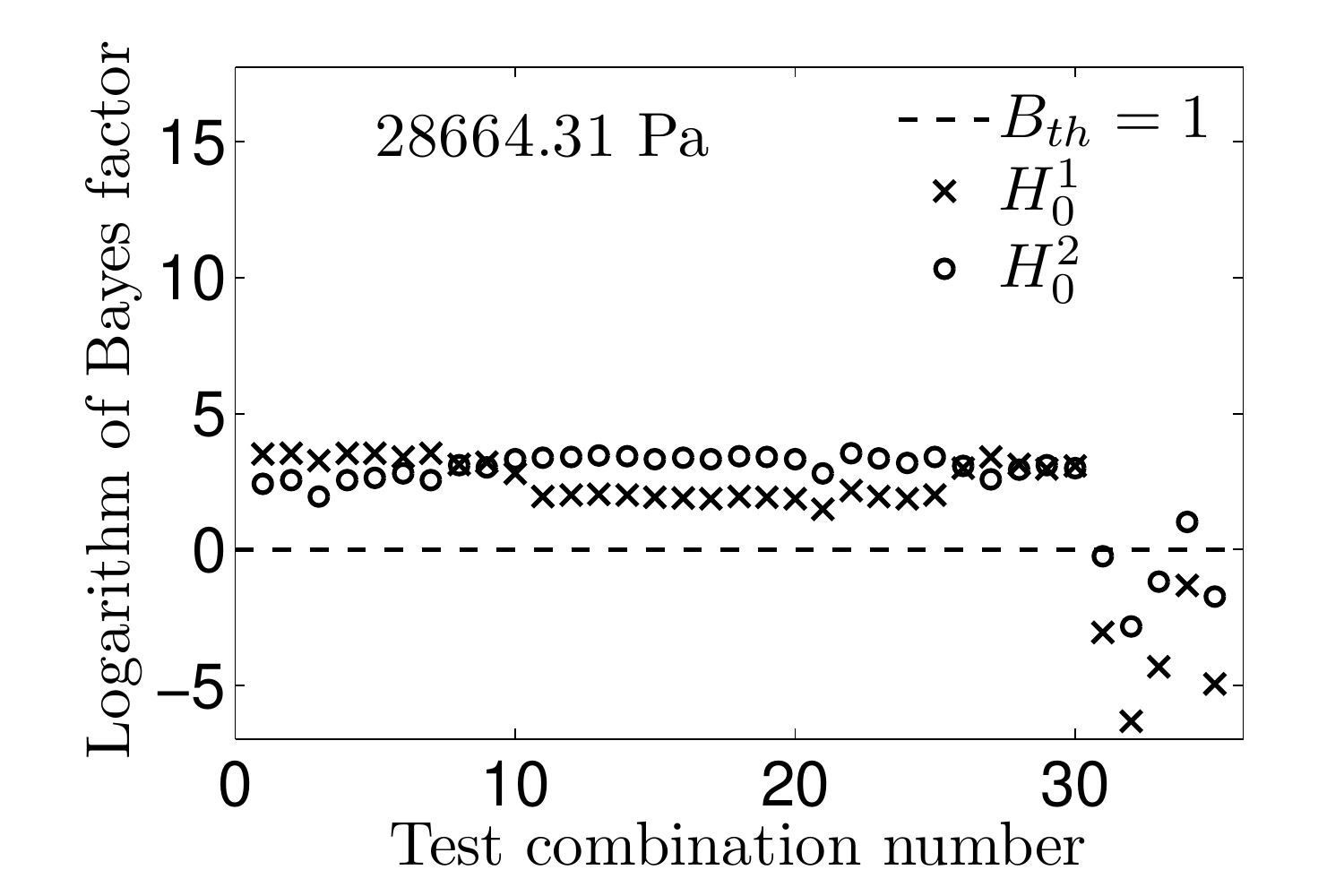}\label{fig:IntBFmetric2}}\\
\subfigure[]{
\includegraphics[width=0.46\textwidth]{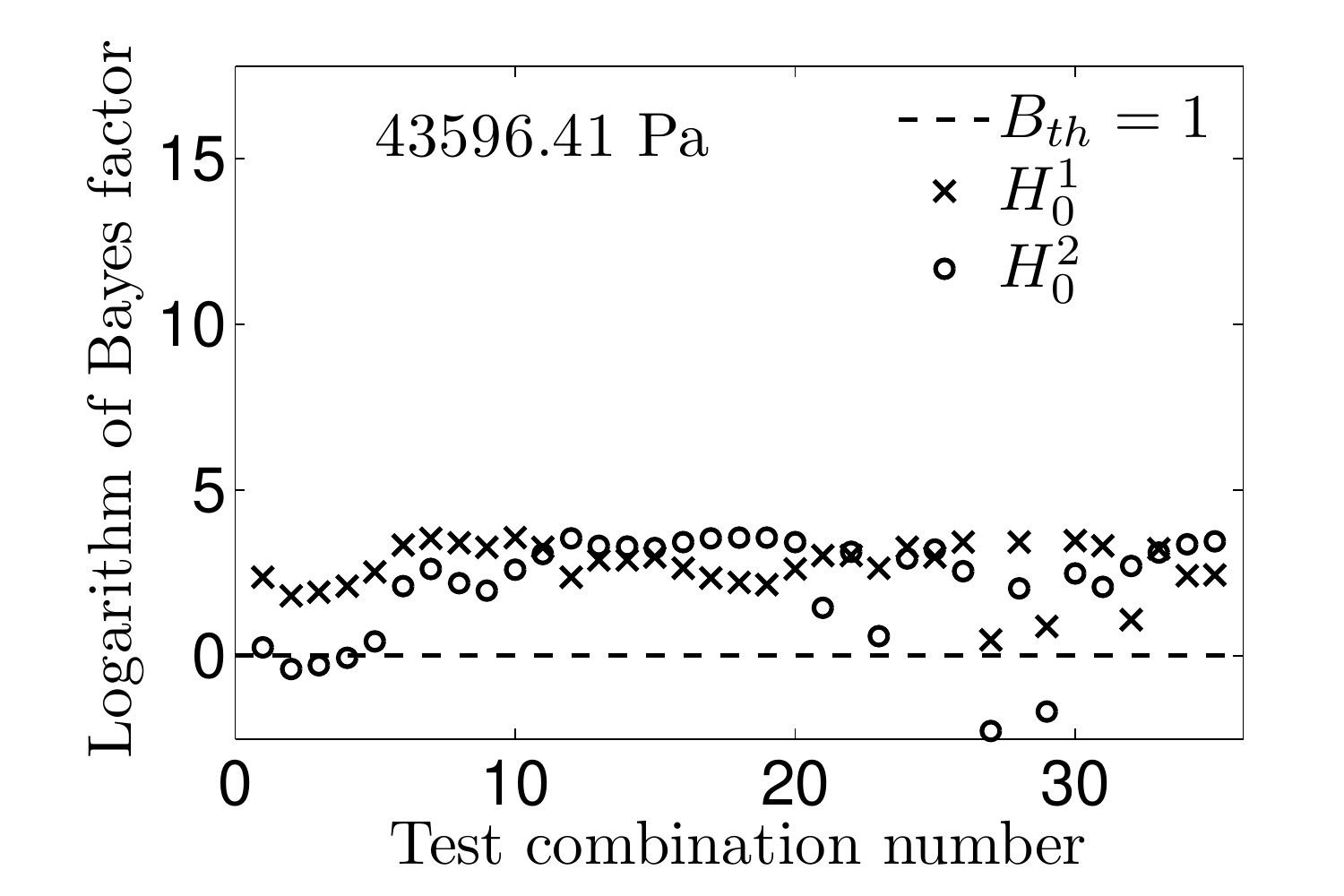}}
\subfigure[]{
\includegraphics[width=0.46\textwidth]{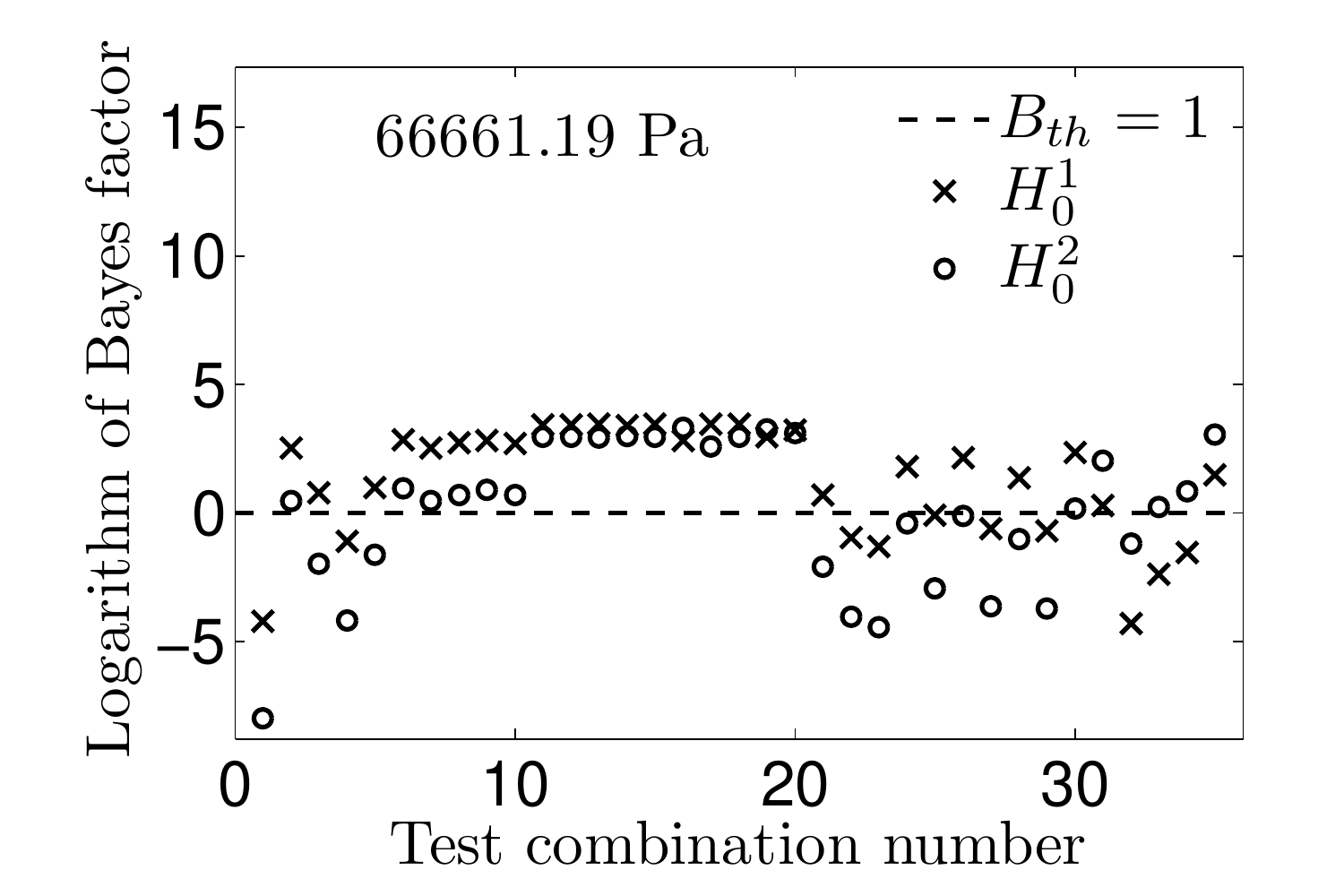}}
\end{center}
\caption{Bayes factor in interval-based hypothesis testing (on logarithmic scale)}
\label{fig:IntBFmetric}
\end{figure}

\paragraph{Interval hypothesis on distribution parameters}
As discussed in Section~\ref{section:BHT}, combination of two Bayesian hypothesis tests based on the interval null hypotheses $H_0^1$ and $H_0^2$ respectively can ref{}lect the existence of directional bias. In practical, the parameters $\epsilon_{\mu}$, $\epsilon_{\sigma1}$, and $\epsilon_{\sigma2}$ that def{}ine the intervals can be determined based on the strictness requirement of validation. For the purpose of illustration, we set $\epsilon_{\mu}=0.025$, $\epsilon_{\sigma1}=-0.005$, and $\epsilon_{\sigma2}=0.005$. $\mu_l$ and $\mu_u$ that def{}ine the possible range of $\mu$ are set as 0 and 1 respectively since the MEMS device considered is under-damped. $\sigma_l$ and $\sigma_u$ are set to be 0.001 and 0.05 respectively. The results of Bayesian interval hypothesis testings are calculated using Eq.~\ref{eq:numIntBHT} -~\ref{eq:LKintBHT1}, and are shown in Fig.~\ref{fig:IntBFmetric} and Table~\ref{table:gPcInIntBHT}.

\begin{table}[h!]
\begin{center}
\caption{Performance of gPC models in interval-based Bayesian hypothesis testing with $\log B_{th} =0$}
\label{table:gPcInIntBHT}
\begin{tabular}{ccccccc}
\toprule
 & & Pressure (Pa) & 18798.45 & 28664.31 & 43596.41 & 66661.19\\
\midrule
\multirow{2}{*}{$H_0^1:$} & $-\epsilon_{\mu} \le \mu_m-\mu \le 0$ & Number of failures & 10 & 5 & 0 & 10\\
 & $\epsilon_{\sigma1} \le |\sigma_m-\sigma| \le \epsilon_{\sigma2}$ & Overall Bayes factor & 3.1 & 58.3 & 92.9 & 44.1\\
\cmidrule(r){3-7}
\multirow{2}{*}{$H_0^2:$} & $0 \le \mu_m-\mu \le \epsilon_{\mu}$ & Number of failures & 5 & 4 & 5 & 14\\
& $\epsilon_{\sigma1} \le |\sigma_m-\sigma| \le \epsilon_{\sigma2}$ & Overall Bayes factor & 63.9 & 87.1 & 74.1 & 1.4\\
\cmidrule(r){3-7}
& \multirow{2}{*}{Combined test} & Number of failure & 10 & 5 & 5 & 16\\
& & Failure percentage & 28.6\% & 14.3\% & 14.3\% & 45.7\%\\
\bottomrule
\end{tabular}
\end{center}
\end{table}

\paragraph{Equality hypothesis on probability density functions}
In this study, the possible values of damping coef{}f{}icient range from 0 to 1 since the system is under-damped. Hence the limit of integration in the denominator of Eq.~\ref{eq:BFvalidation} is [0, 1], while the limit of integration in the numerator is $[-\infty,+\infty]$.

\begin{figure}[h!]
\begin{center}
\subfigure[]{
\includegraphics[width=0.46\textwidth]{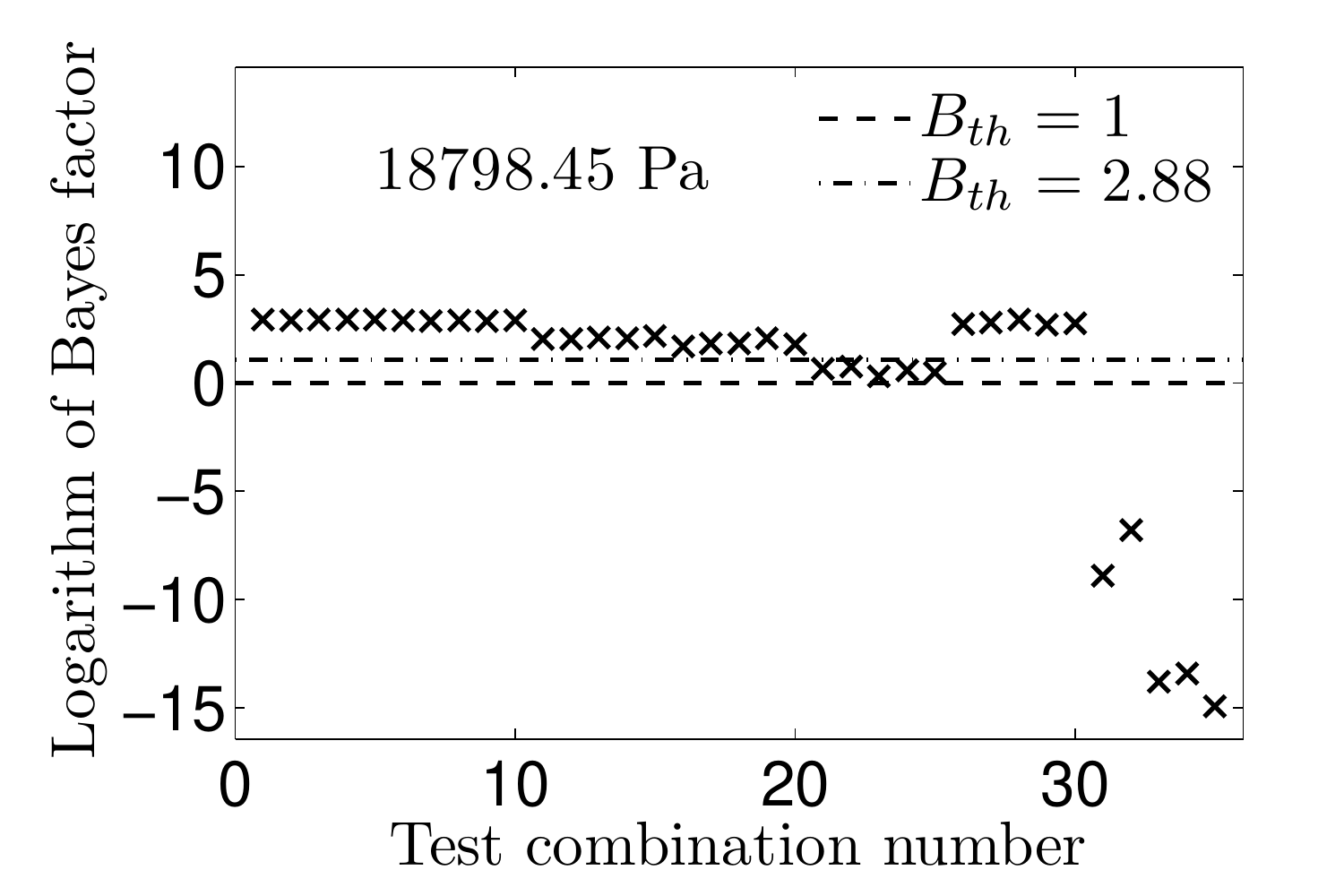}}
\subfigure[]{
\includegraphics[width=0.46\textwidth]{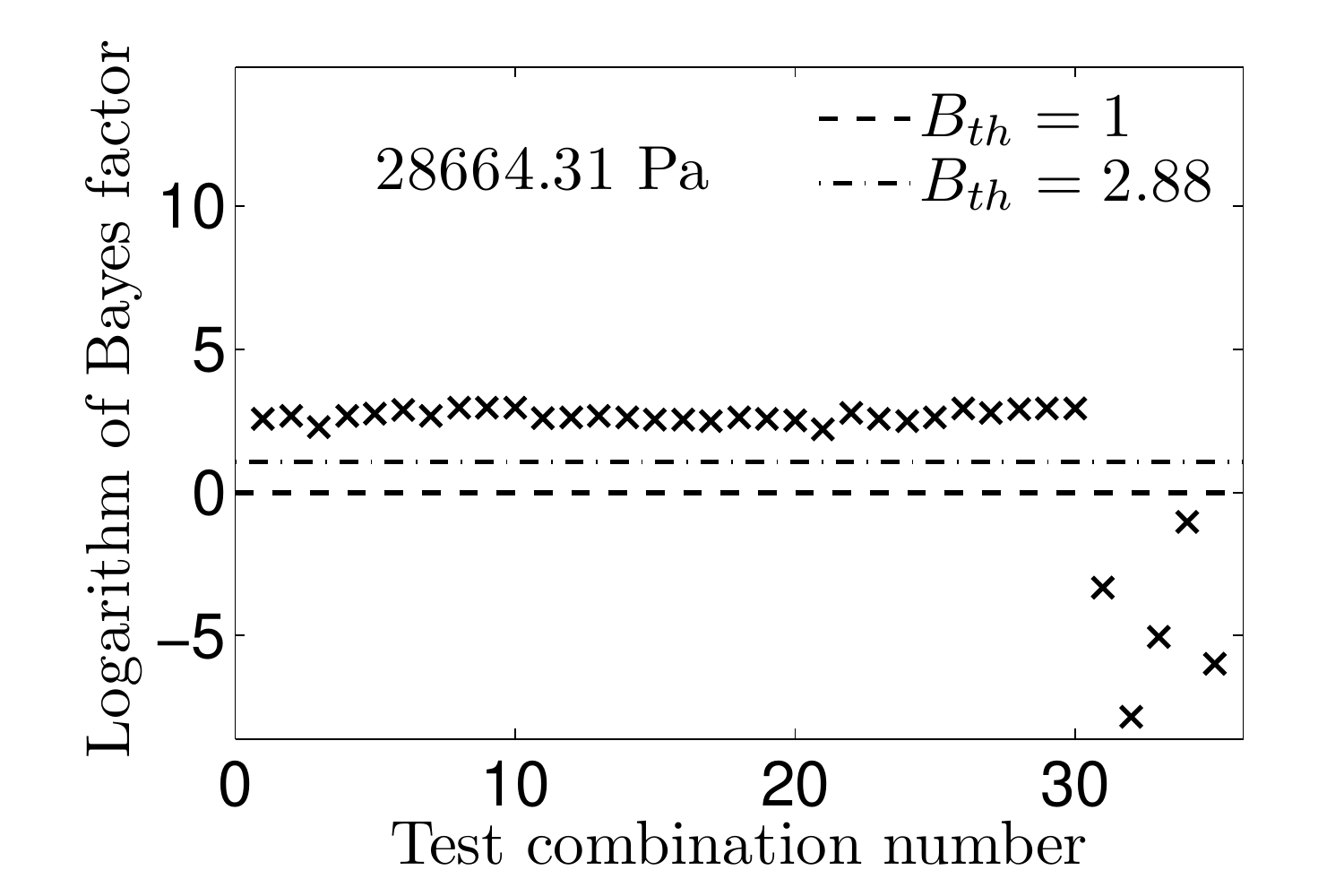}\label{fig:BFmetric2}}\\
\subfigure[]{
\includegraphics[width=0.46\textwidth]{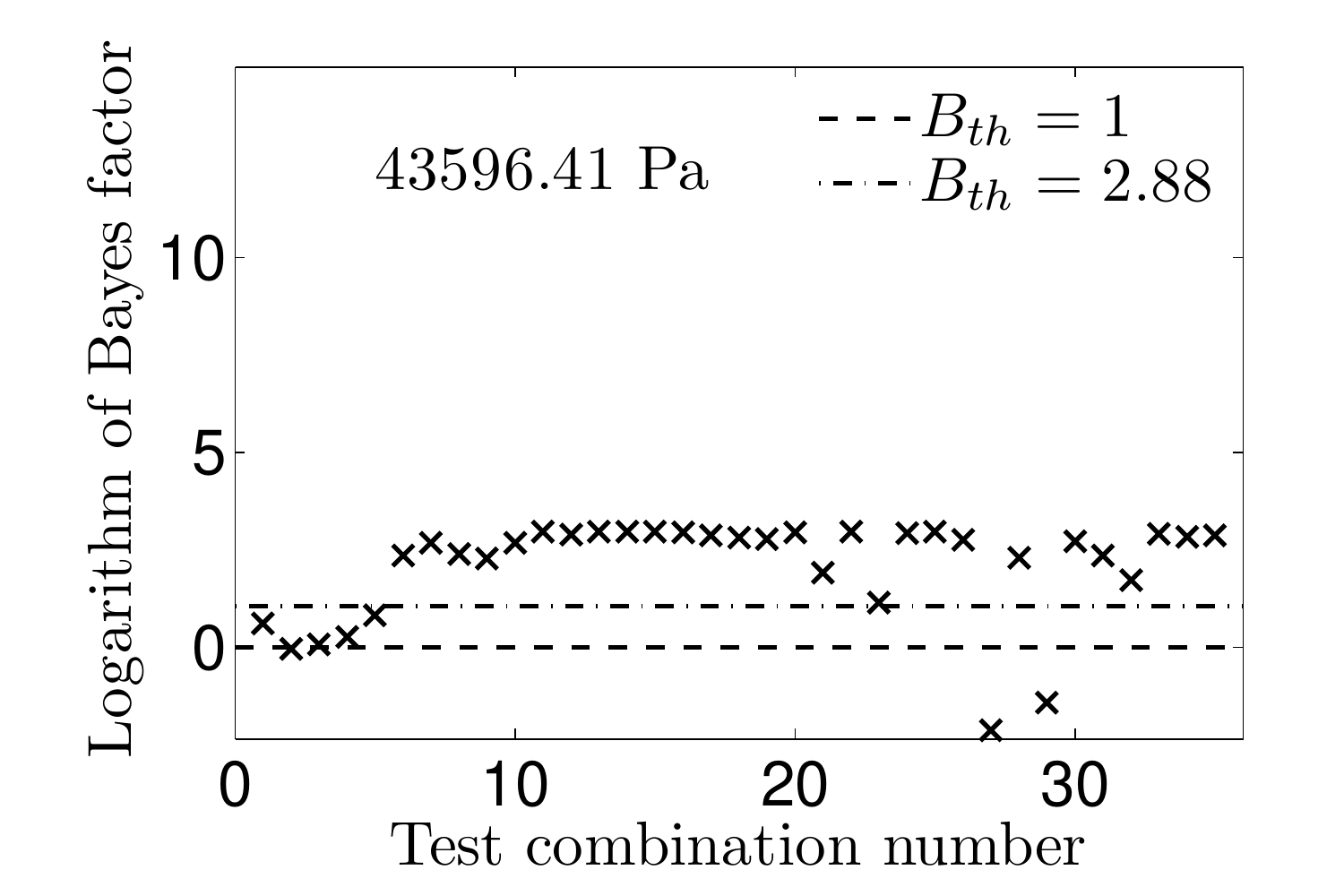}}
\subfigure[]{
\includegraphics[width=0.46\textwidth]{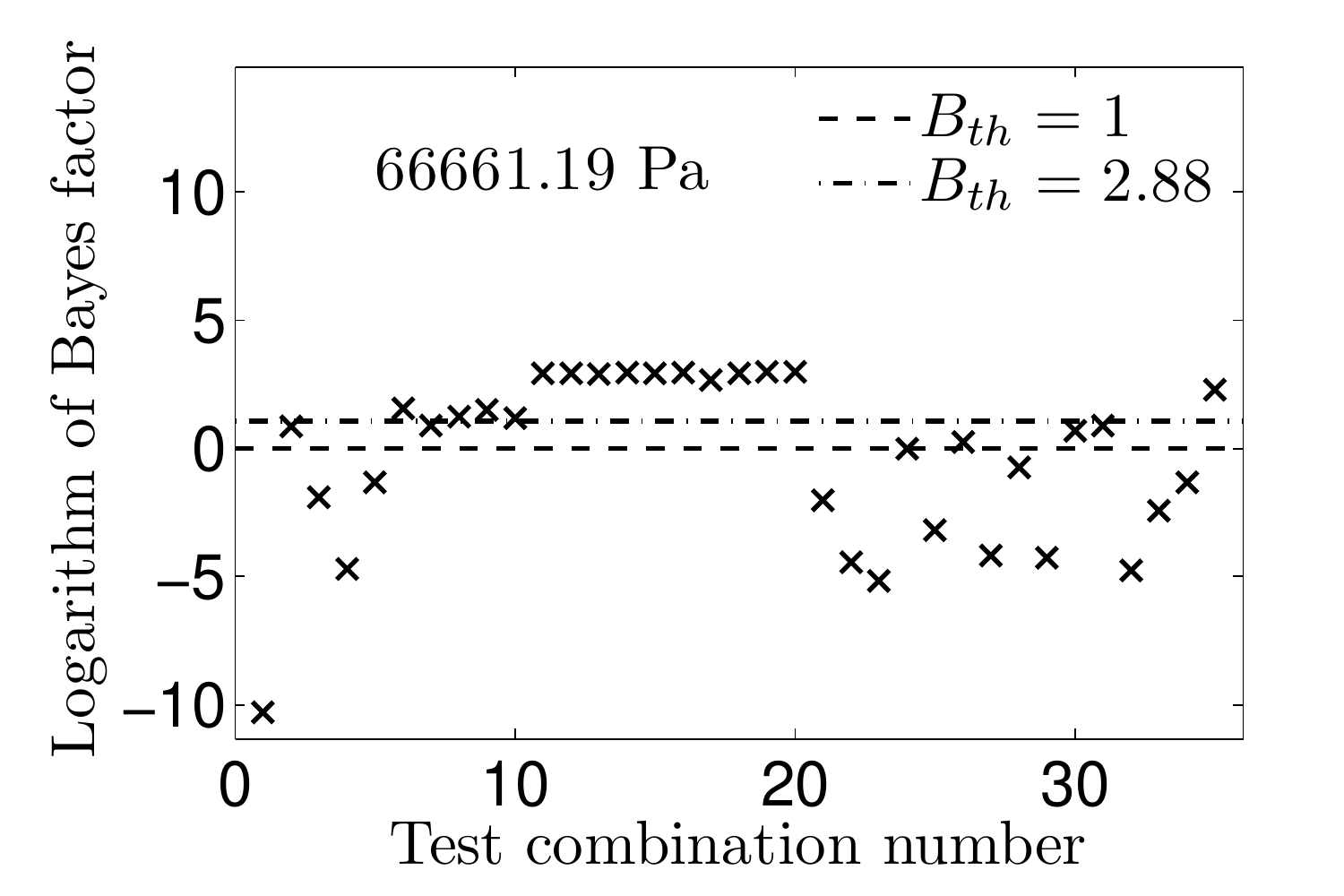}}
\end{center}
\caption{Bayes factor in equality-based hypothesis testing (on logarithmic scale)}
\label{fig:BFmetric}
\end{figure}

The performance of the gPC models in Bayesian hypothesis testing are shown in Fig.~\ref{fig:BFmetric} and Table~\ref{table:gPcInBHT}. The values of Bayes factor are calculated using Eq.~\ref{eq:BFvalidation}, and the threshold Bayes factor $B_{th} = 1$ (this threshold value is chosen based on the discussion in Section~\ref{section:BHT}). Although the performance of the gPC model in terms of failure percentage is dif{}ferent for the two hypothesis tests as shown in Table~\ref{table:gPcInztest} and Table~\ref{table:gPcInBHT}, if one increases the threshold Bayes factor $B_{th}$ to 2.88, which is calculated using Eq.~\ref{eq:BFandpvalueInztest} with $p = 0.05$ in Section~\ref{section:relationBFandpvalue}, the result of Bayesian hypothesis testing in terms of the number of failures becomes the same as in the $z$-test in Section~\ref{section:numExpCHT}. The reason for this coincidence has been explained in Section~\ref{section:relationBFandpvalue}. Note that the performance of the second gPC model (for pressure = 28664.31 Pa) remains the same when $B_{th}$ is raised from 1 to 2.88, and this can be easily observed from Fig.~\ref{fig:BFmetric2}.

\begin{table}[h!]
\begin{center}
\caption{Performance of gPC models in equality-based hypothesis testing with $\log B_{th} =0$}
\label{table:gPcInBHT}
\begin{tabular}{lcccc}
\toprule
Pressure (Pa) & 18798.45 & 28664.31 & 43596.41 & 66661.19\\
\midrule
Number of failures & 5 & 5 & 3 & 15\\
Failure percentage & 14.3\% & 14.3\% & 8.6\% & 42.9\%\\
Overall Bayes factor (log-scale) & 7.4 & 57.2 & 72.3 & -10.2\\
\bottomrule
\end{tabular}
\end{center}
\end{table}

By comparing the results based on interval hypothesis on distribution parameters and equality hypothesis on probability density functions (Tables~\ref{table:gPcInIntBHT} and~\ref{table:gPcInBHT}), it can be observed that the performance of the gPC model for pressure 18798.45 Pa in the f{}irst test is signif{}icantly worse than in the second test, while the models for the other three pressures have similar failure percentages in these two tests.  As shown in Fig.~\ref{fig:GraphP1}, the data are all located below the mean predictions of this gPC model, which indicates the existence of directional bias, and thus the gPC model for pressure 18798.45 Pa performs worse in the Bayesian interval hypothesis testing.
\subsection{Validation using non-hypothesis testing-based methods}
\subsubsection{Reliability-based metric}
\begin{figure}[h!]
\begin{center}
\subfigure[]{
\includegraphics[width=0.46\textwidth]{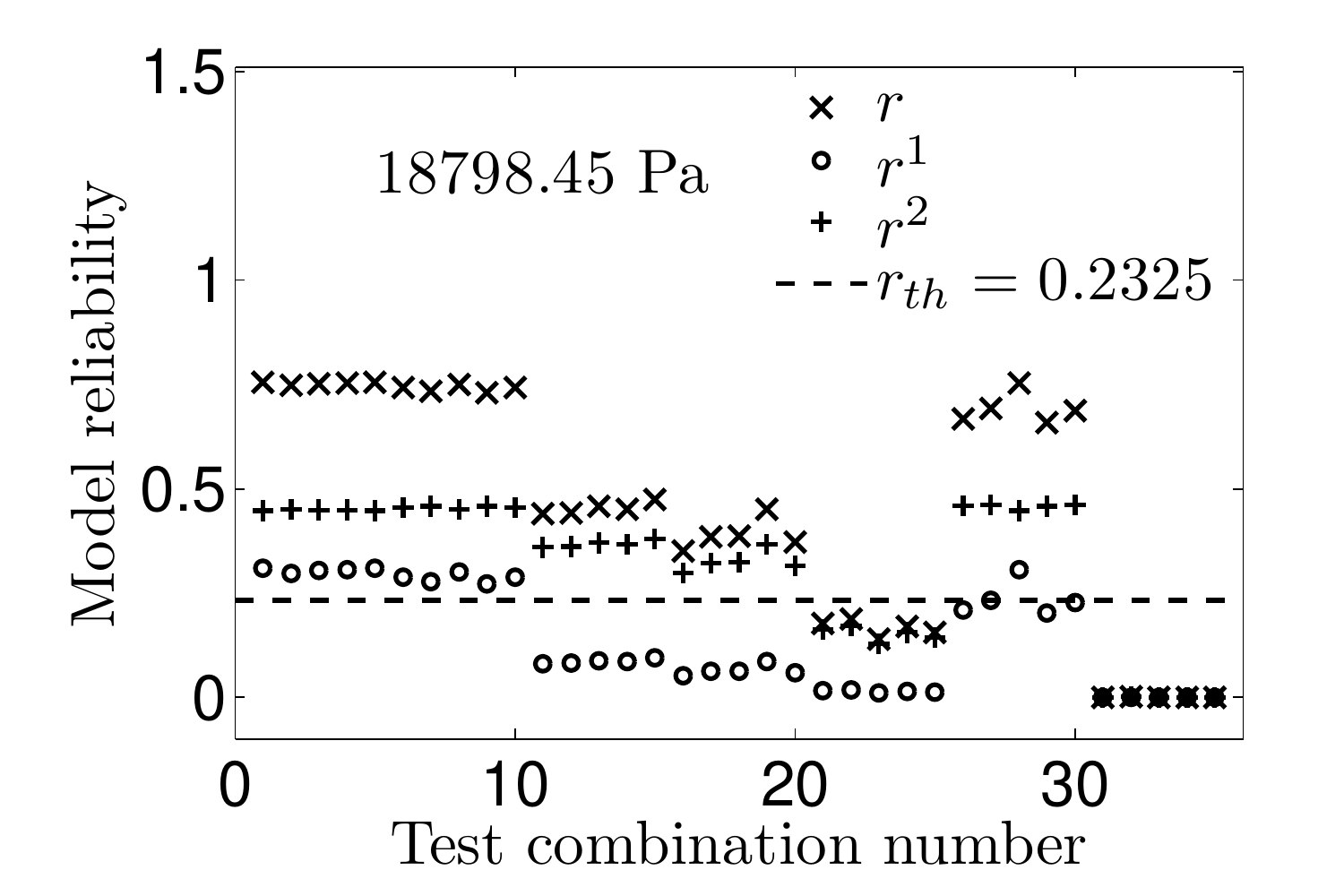}}
\subfigure[]{
\includegraphics[width=0.46\textwidth]{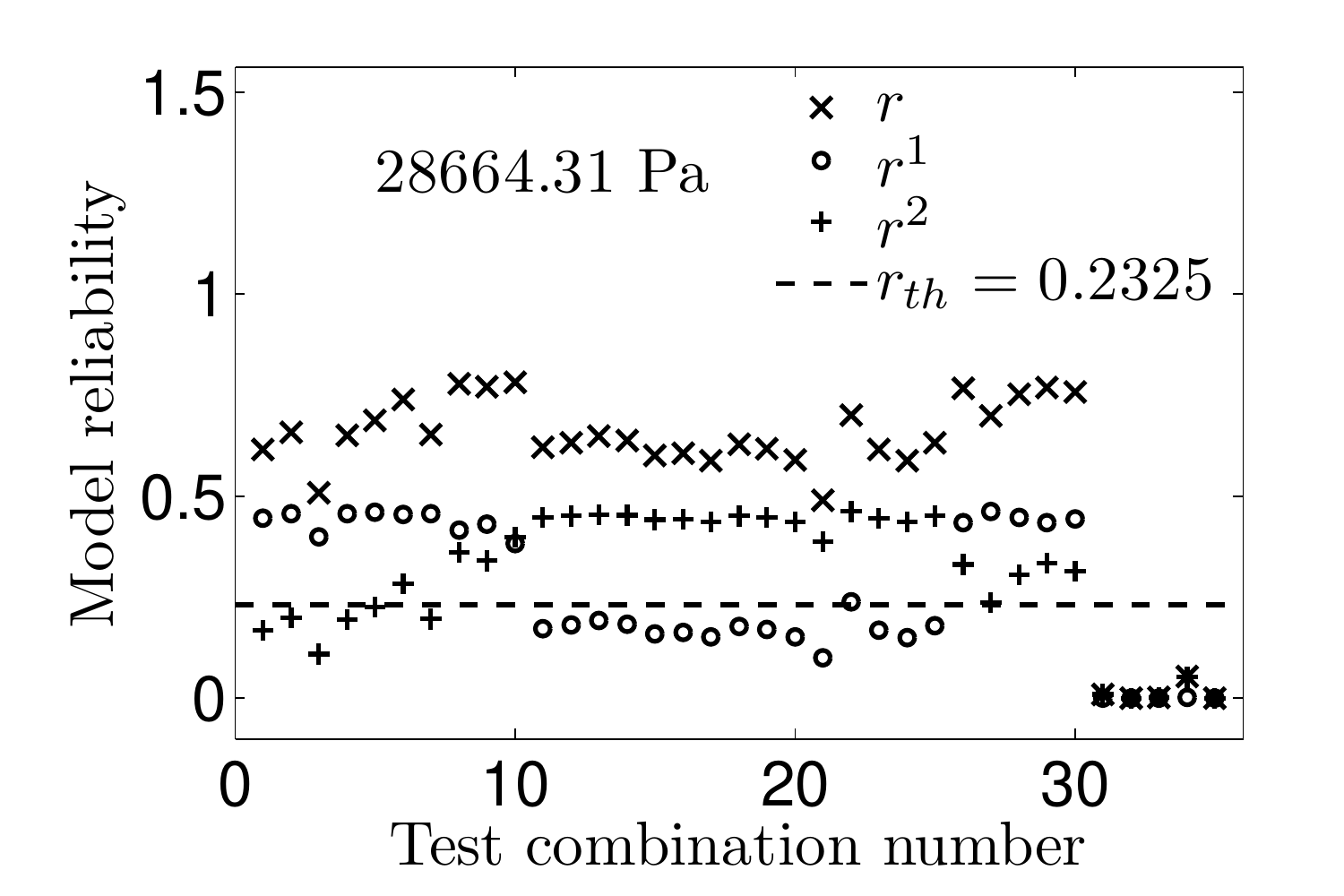}}\\
\subfigure[]{
\includegraphics[width=0.46\textwidth]{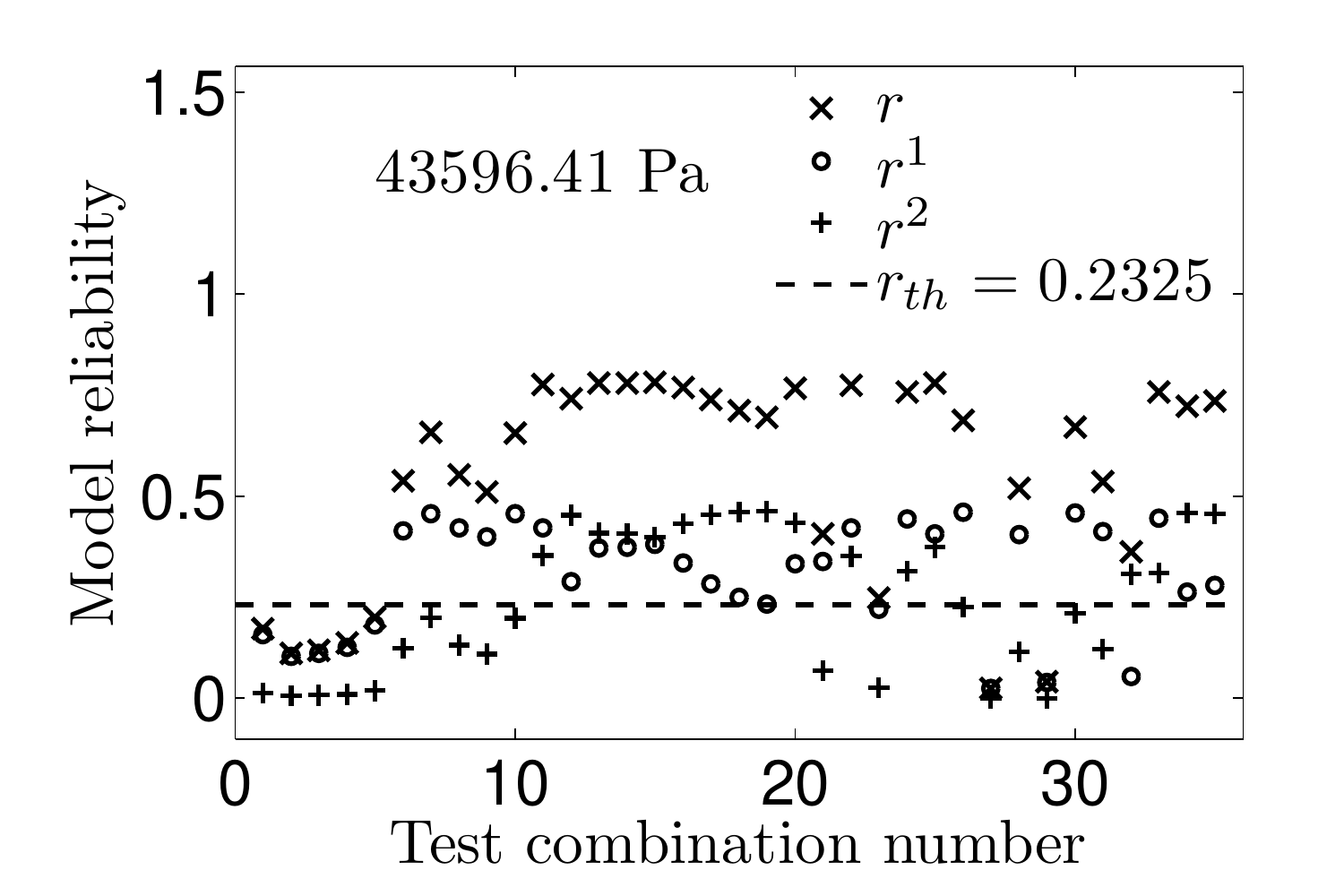}}
\subfigure[]{
\includegraphics[width=0.46\textwidth]{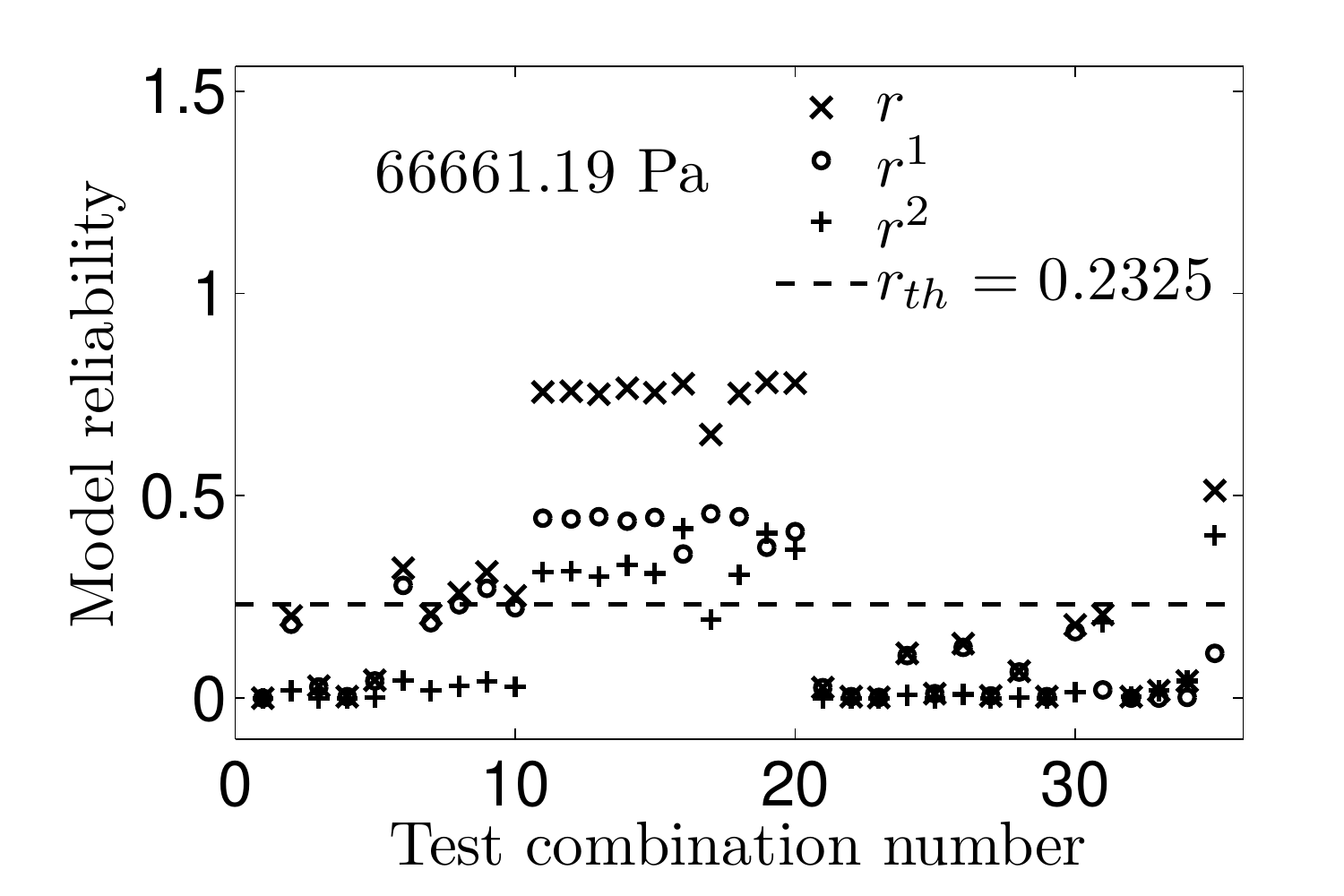}}
\end{center}
\caption{Reliability-based metric}
\label{fig:MRmetric}
\end{figure}

Fig.~\ref{fig:MRmetric} and Table \ref{table:gPcInMRA} show the calculated values of the reliability-based metric $r$, $r^1$ and $r^2$ (Eq.~\ref{eq:rmformula} and~\ref{eq:rmformula1}), the failure percentage of models with $\epsilon = 0.025$ and the decision criterion $r_{th} = 0.2325$. This decision criterion is obtained using Eq.~\ref{eq:rmAndp} with the signif{}icance level $\alpha = 0.05$, and thus the results of validation (comparing $r$ with $r_{th}$) in terms of failure percentage are the same as in the $z$-test in Section~\ref{section:numExpCHT}. It can also observed that the failure percentage of the gPC model for pressure 18798.45 Pa increases signif{}icantly in the test that comparing $r^1$ and $r^2$ with $r_{th}/2$ due to the existence of directional bias.

\begin{table}[h!]
\begin{center}
\caption{Performance of gPC models in reliability-based method with $r_{th}=0.69$}
\label{table:gPcInMRA}
\begin{tabular}{lccccc}
\toprule
& Pressure (Pa) & 18798.45 & 28664.31 & 43596.41 & 66661.19\\
\midrule
\multirow{2}{*}{$r$ vs. $r_{th}$} & Number of failures & 10 & 5 & 7 & 20\\
& Failure percentage & 28.6\% & 14.3\% & 20.0\% & 57.1\%\\
\cmidrule(r){2-6}
\multirow{2}{*}{$r^1$ and $r^2$ vs. $r_{th}/2$} & Number of failures & 20 & 7 & 12 & 25\\
& Failure percentage & 57.1\% & 20.0\% & 34.3\% & 71.4\%\\
\bottomrule
\end{tabular}
\end{center}
\end{table}
\subsubsection{Area metric-based method}
\begin{figure}[h!]
\begin{center}
\subfigure[]{
\includegraphics[width=0.46\textwidth]{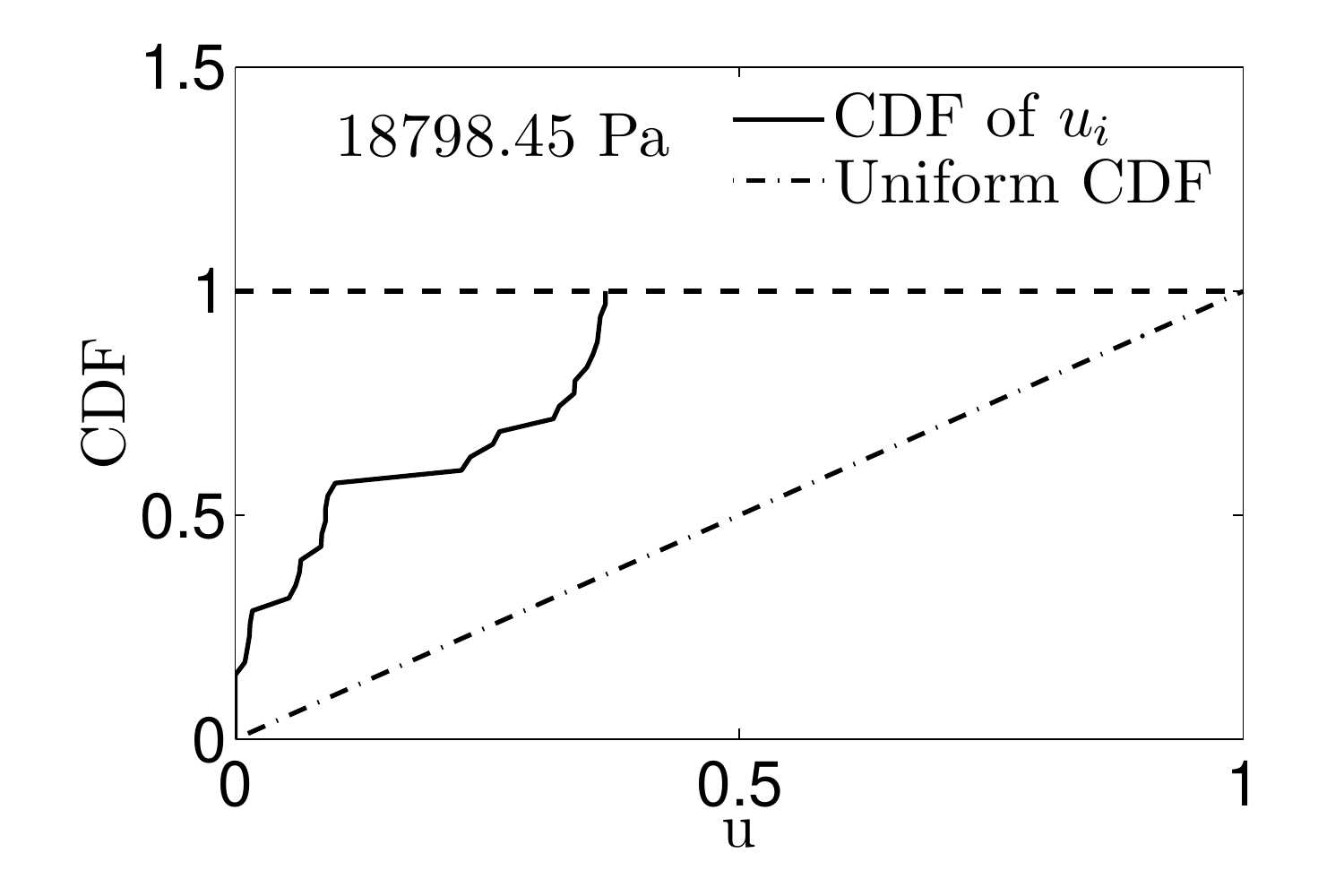}}
\subfigure[]{
\includegraphics[width=0.46\textwidth]{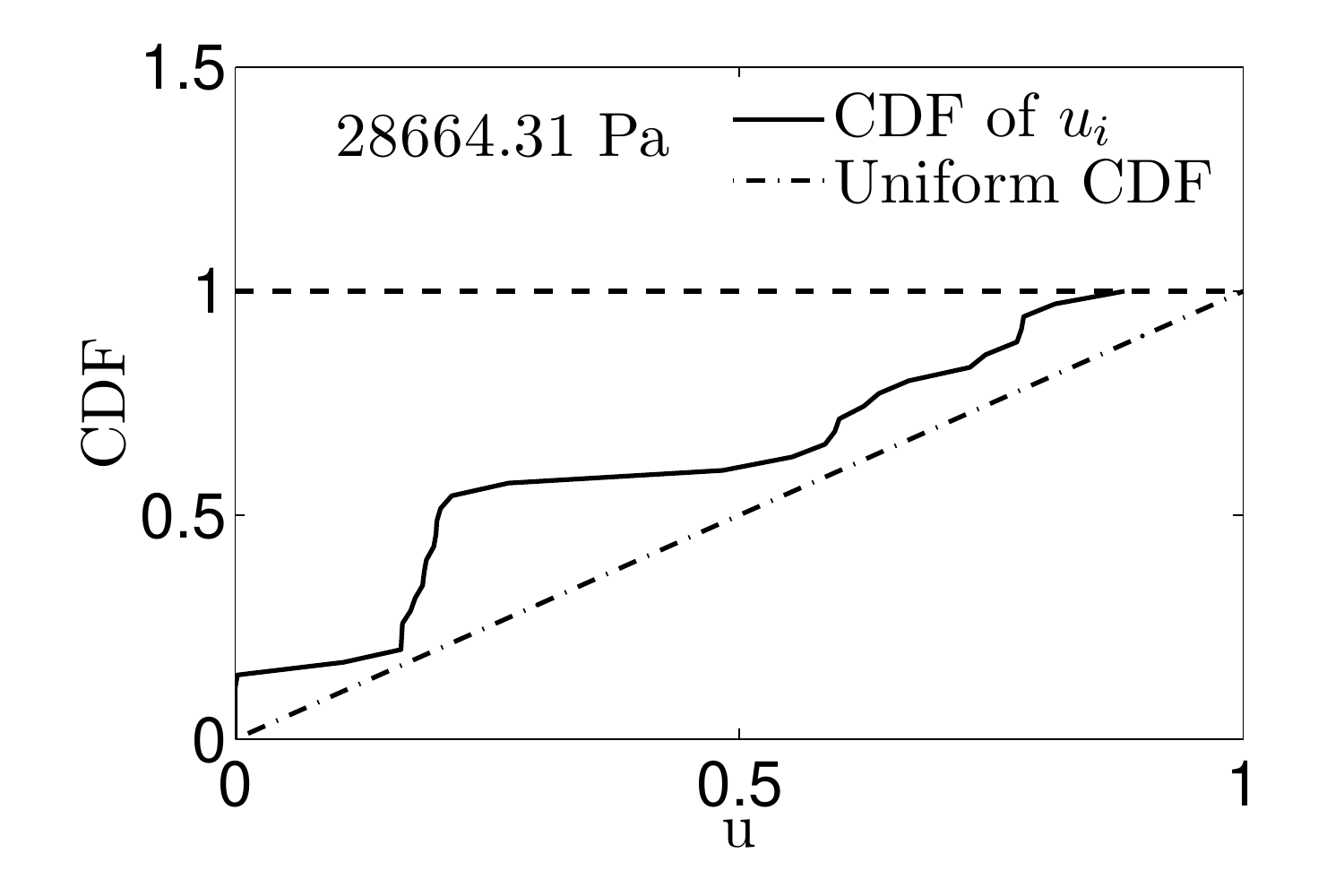}}\\
\subfigure[]{
\includegraphics[width=0.46\textwidth]{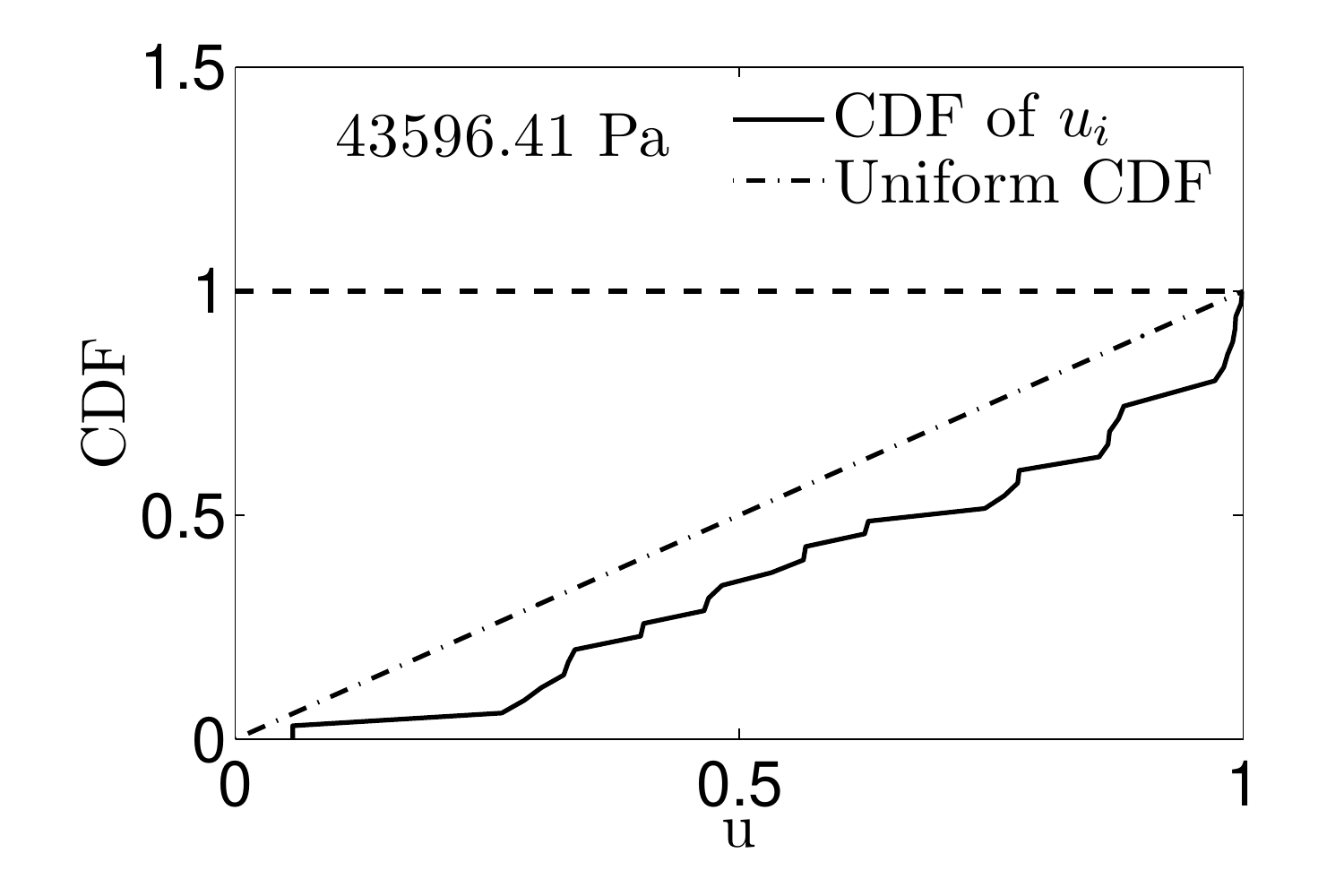}}
\subfigure[]{
\includegraphics[width=0.46\textwidth]{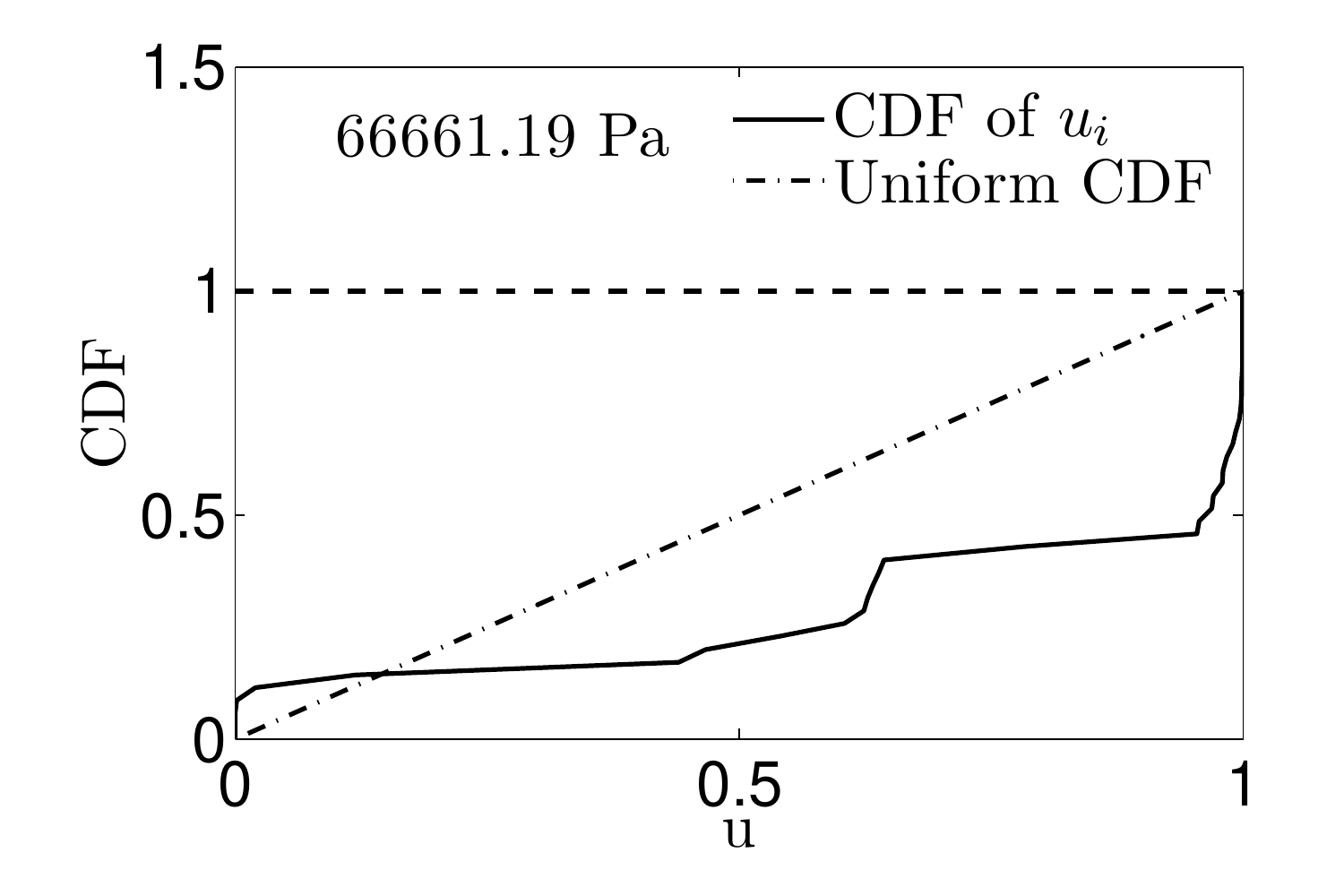}}
\end{center}
\caption{Empirical CDF of $u_i$ and standard uniform CDF}
\label{fig:ARmetric}
\end{figure}

The area metric for the four gPC models can be computed using Eqs.~\ref{eq:upooling} and~\ref{eq:areametric}, and the results are shown in Fig.~\ref{fig:ARmetric} and Table~\ref{table:gPcInam}. Note that the gPC model for pressure 18798.45 Pa has the highest area value and thus performs worst. This is due to the directional bias between mean predictions and experimental data, and it is ref{}lected in the area metric as discussed in Section~\ref{section:areaMetric}.

\begin{table}[h!]
\begin{center}
\caption{Area metric for gPC models}
\label{table:gPcInam}
\begin{tabular}{lcccc}
\toprule
Pressure (Pa) & 18798.45 & 28664.31 & 43596.41 & 66661.19\\
\midrule
$d(F_u,S_u)$ & 0.543 & 0.146 & 0.151 & 0.250\\
\bottomrule
\end{tabular}
\end{center}
\end{table}
\subsection{Discussion}
This section demonstrated a numerical example of validating the gPC surrogate model for the RF switch damping coef{}f{}icient using the validation methods presented in Sections~\ref{section:binHT} and~\ref{section:nonHT}, and 140 fully characterized experimental data points. Based on the performance of the gPC model in these validation tests, it can be concluded that the prediction of the gPC model has better agreement with observation under the middle two values of pressure (28664.31 Pa and 43596.41 Pa), whereas less agreement can be found under the lowest and highest pressure values (18798.45 Pa and 66661.19 Pa).  The decision on model acceptance can be formed based on the failure percentages with the hypothesis testing methods and the reliability-based method, and the values of area-based metric, given the desired acceptance threshold. It is shown that the $z$-test and the reliability-based metric give the same results in terms of failure percentage when $r_{th}$ is selected based on the signif{}icance level $\alpha$ used in $z$-test. Similarly, classical and Bayesian hypothesis testing give the same result by choosing a specif{}ic threshold Bayes factor as shown in Section~\ref{section:relationBFandpvalue}. It is also shown that the existence of directional bias can be ref{}lected in the Bayesian interval hypothesis testing, reliability-based method with modif{}ied intervals, and the area metric-based method. Models that have directional bias will perform worse in these three validation methods than in classical hypothesis testing and in Bayesian hypothesis testing with equality hypothesis on probability density functions.
\section{Conclusion}
This paper explored various quantitative validation methods, including classical hypothesis testing, Bayesian hypothesis testing, a reliability-based method, and an area metric-based method, in order to validate computational model prediction. The numerical example featured a generalized polynomial chaos (gPC) surrogate model which predicts the micro-scale squeeze-f{}ilm damping coef{}f{}icient for RF MEMS switches. 

An Bayesian interval hypothesis testing-based method is formulated, which validates the accuracy of the predicted mean and standard deviation from a model, taking into account the existence of directional bias. Further, Bayesian hypothesis testing to validate the entire PDF of model prediction is formulated. These two formulations of Bayesian hypothesis testing can be applied to both fully characterized and partially characterized experiments, and the case when multiple validation points are available. It is shown that while the classical hypothesis testing is subject to type I and type II error, the Bayesian hypothesis testing can minimize such risk by (1) selecting a risk-based threshold, and (2) subsequent model averaging using posterior probabilities. It is observed that under some conditions, the $p$-value in the $z$-test or $t$-test can be mathematically related to the Bayes factor and the reliability-based metric. 

The area metric is also sensitive to the direction of bias between model predictions and experimental data, and so is the reliability-based method. The Bayesian model validation result and reliability-based metric can be directly incorporated in long-term failure and reliability analysis of the device, thus explicitly accounting for model uncertainty, whereas the area-based metric lacks a direct interpretation for its results. In addition, due to the use of likelihood function in the Bayesian hypothesis testing, the Bayesian model validation method can be extended to the case that the validation data is in the form of interval, as shown in~\cite{Sankararaman2011a,Sankararaman2011b}.
\section*{Acknowledgment}
This paper is based upon research partly supported by the Department of Energy [National Nuclear Security Administration] under Award Number DE-FC52-08NA28617 to Purdue University (Principal Investigator: Prof. Jayathi Murthy), and Subaward to Vanderbilt University. The support is gratefully acknowledged. The authors also thank the U. S. DOE (NNSA) PSAAP Center for Prediction of Reliability, Integrity and Survivability of Microsystems (PRISM) at Purdue University for providing the models and validation data for the numerical example.
\bibliographystyle{elsarticle-num}
\bibliography{Research}

\end{document}